\documentclass[aps, prb, preprint, superscriptaddress, endfloats*]{revtex4-2}
\usepackage{amsmath,hyperref,cleveref,amssymb,amsfonts,graphicx,multirow,color,bm,textcomp,mathtools}
\usepackage{paralist}
\usepackage{srcltx}
\usepackage[all,cmtip]{xy}
\usepackage{mathrsfs}
\usepackage{srcltx,soul,subfigure}
\usepackage{threeparttable,dsfont,bbm}
\usepackage{comment}
\usepackage{dcolumn}

\usepackage{xcolor}

\begin{document}
\title{Symmetry-Enforced Dirac Fermions and Structural Metastability in Pentagonal Monolayers of Transition-Metal Ditellurides}

\author{Manoj Gadtoula}
\affiliation {Department of Physics and Astronomy, University of Missouri, Columbia, Missouri 65211, USA}
\author{Clayton Conner}
\affiliation {Department of Physics and Astronomy, University of Missouri, Columbia, Missouri 65211, USA}
\author{Avinash Sah}
\affiliation {Department of Physics and Astronomy, University of Missouri, Columbia, Missouri 65211, USA}
\author{Ronghao Luo}
\affiliation {Department of Physics and Astronomy, University of Missouri, Columbia, Missouri 65211, USA}
\author{Guang~Bian}\email{biang@missouri.edu}
\affiliation {Department of Physics and Astronomy, University of Missouri, Columbia, Missouri 65211, USA}
\affiliation {MU Materials Science \& Engineering Institute, University of Missouri, Columbia, Missouri 65211, USA}

\newpage
\begin{abstract}
The recent synthesis of pentagonal PdTe$_2$ monolayer motivates broader research interests in transition-metal ditellurides whose electronic phases are governed by symmetry and structural reconstruction. The pentagonal phase of transition-metal ditellurides exhibits electronic properties that are dramatically different from those of its hexagonal counterpart due to its lower crystalline symmetry. Using first-principles calculations, we study monolayer $X$Te$_2$ ($X=\mathrm{Pd},\mathrm{Pt},\mathrm{Ni}$) in both hexagonal and pentagonal polymorphs. By constructing a continuous structural interpolation between the hexagonal and pentagonal phases, we show that the semimetal (hex)-to-semiconductor (penta) transition occurs only after an intermediate structural threshold rather than at the onset of symmetry reduction. The gap opening coincides with the formation of Te--Te dimers, which drive the bonding--antibonding splitting of the Te $p$ states and reorganize the band edges. In addition, the nonsymmorphic symmetry of the pentagonal phase enforces band degeneracies at the Brillouin-zone boundary, leading to symmetry-protected two-dimensional (2D) Dirac states. These results establish pentagonal $X$Te$_2$ monolayers as a new class of 2D semiconductors in which symmetry constraints and local bonding collectively shape the unconventional semiconducting electronic structure.

\end{abstract}


\maketitle

\newpage

\section{Introduction}

Two-dimensional transition-metal dichalcogenides (TMDs) provide a versatile platform for studying how crystal symmetry, dimensionality, and spin--orbit coupling (SOC) shape electronic structure, transport, and topology \cite{Wang2012,Chhowalla2013,Manzeli2017}. Within this family, the same chemical composition can adopt distinct lattice geometries with different electronic properties, enabling phase engineering between semiconducting, metallic, and semimetallic states \cite{Duerloo2014,Li2016,Wang2017,Cho2015}. This structural tunability has led to phase-controlled functionality in layered materials, including topological phases in distorted TMD polymorphs \cite{Qian2014} and other emergent topological states in intercalated TMD systems \cite{Sah2026}. The group-10 ditellurides PdTe$_2$, PtTe$_2$, and NiTe$_2$ are especially interesting in this context because their well-known hexagonal 1T-derived layered phases are closely connected to topological semimetal physics. In bulk and multilayer forms, these compounds have been identified as hosting type-II Dirac fermions and related topological surface states \cite{Huang2016,Bahramy2018,Noh2017,Yan2017,Xu2018}. The type-II classification, originally introduced in the context of Weyl semimetals \cite{Soluyanov2015}, provides the broader framework for understanding these strongly tilted Dirac dispersions. This establishes the $X$Te$_2$ ($X=\mathrm{Pd},\mathrm{Pt},\mathrm{Ni}$) family as a promising platform for exploring how structural reconstruction can qualitatively alter band topology and Dirac dispersions. Among the structural motifs accessible to two-dimensional ditellurides, the pentagonal phase represents a particularly dramatic departure from the parent hexagonal lattice, one that has only recently been realized in experiment.

Pentagon-based two-dimensional materials have emerged as a distinct structural class, exhibiting properties that differ fundamentally from those of conventional hexagonal lattices \cite{Shen2022,Liang2022,Zhang2015}. 
Following the prediction of penta-graphene \cite{Zhang2015}, a series of theoretical and experimental studies have established pentagonal monolayers as realistic candidates across multiple chemistries, including noble-metal chalcogenides \cite{Oyedele2017,Kempt2020} and transition-metal disulfides \cite{Lv2021}. Symmetry-based analyses have further predicted topological band features in pentagonal lattices. These include Weyl nodes pinned at high-symmetry points and nodal lines along the Brillouin-zone boundary in noncentrosymmetric pentagonal monolayers \cite{Bravo2021}, as well as topologically nontrivial conduction bands in pentagonal PdSe$_2$ \cite{Bravo2022}. In particular, pentagonal PdTe$_2$ was recently synthesized as a metastable two-dimensional phase by symmetry-driven epitaxy \cite{Liu2024}. Since its synthesis, pentagonal PdTe$_2$ has spurred extensive theoretical studies exploring defect-engineered electronic and magnetic properties \cite{Sharma2024}, strain-tunable nonlinear optical responses \cite{Sun2025}, bulk photovoltaic effects in stacked bilayers \cite{Hou2025}, potential sensing applications \cite{Han2025}, and hydrogen evolution catalysis \cite{Parkar2026}. These recent findings build upon earlier predictions of anisotropic electronic and thermoelectric responses in pentagonal Pd- and Pt-based chalcogenides \cite{Lan2019,Tao2021,Xiong2019}, further highlighting the versatile functionality of this material class.

The pentagonal crystal structure is associated with reduced symmetry and nonsymmorphic symmetry operations, which can enforce robust band degeneracies that are unavailable in symmorphic crystals. It is known that nonsymmorphic symmetries such as glide mirrors and screw axes give rise to Dirac crossings, nodal structures, and other symmetry-enforced semimetallic features in both two- and three-dimensional systems \cite{Young2015,Fang2015,Wieder2016,Zhao2016,Wang2016,Wieder2018}. These ideas have since been demonstrated in various nonsymmorphic materials ranging from ZrSiS to two-dimensional and quasi-two-dimensional Dirac systems \cite{Schoop2016,Guan2017,Kowalczyk2020,Lu2021}. For pentagonal $X$Te$_2$, the relationship between crystal structure and band degeneracy is particularly relevant, as these monolayer systems belong to a nonsymmorphic layer group ($p2_1/b11$). Nonsymmorphic symmetries enforce symmetry-protected band degeneracies even in the presence of SOC \cite{Bradlyn2017,Po2017,Cano2018}. However, how inversion symmetry, acting together with time reversal and the nonsymmorphic glide, shapes the Dirac crossings of the centrosymmetric pentagonal ditellurides has not been established. While nonsymmorphic symmetry dictates the band degeneracies at the Brillouin-zone boundary, the formation of the semiconducting gap in the pentagonal phase is governed by a different mechanism that is rooted in local chemical bonding rather than global symmetry.

 Dimerization-driven electronic reconstruction is a prominent electronic feature in telluride materials. In IrTe$_2$ and related systems, Te--Te and metal--metal bond rearrangements have been shown to produce large changes in electronic structure through bonding--antibonding splitting and Fermi-surface reconstruction \cite{Oh2013,Cao2013,Pascut2014,Mazumdar2015}. In monolayer IrTe$_2$, dimer formation can even stabilize a large-gap insulating ground state \cite{Hwang2022}, highlighting how local bonding instabilities in tellurides can generate new electronic phases. For pentagonal $X$Te$_2$, the Te--Te dimerization can generate a semiconducting gap in a similar way. However, the detailed mechanism of gap opening driven by Te–Te dimerization and the accompanying symmetry reduction remains to be elucidated.

In this work, we systematically investigate the electronic and structural properties of monolayer PdTe$_2$, PtTe$_2$, and NiTe$_2$ in both hexagonal (1T) and pentagonal polymorphs using first-principles calculations. We show that the pentagonal phases are dynamically metastable and, unlike their hexagonal counterparts, exhibit semiconducting band structures while simultaneously hosting symmetry-enforced Dirac band dispersions. For the pentagonal monolayers (layer group $p2_1/b11$, space group $P2_1/c$), we demonstrate that the combined action of inversion, time-reversal, and nonsymmorphic glide symmetry enforces fourfold Dirac degeneracies at the X and Y points of the Brillouin zone in the presence of SOC. We further construct a continuous structural interpolation pathway connecting the hexagonal and pentagonal phases and demonstrate that the semimetal-to-semiconductor transition occurs only after a critical threshold of structural deformation is reached.  By analyzing the energy landscape, bond-length evolution, electron localization functions, charge densities, and atom-projected densities of states along this pathway, we identify Te--Te dimer formation, rather than epitaxial strain, as the microscopic driver of the gap opening. These results reveal distinct roles for symmetry and local bonding: nonsymmorphic symmetry determines where the band degeneracies occur, whereas Te--Te dimer formation drives the opening of the semiconducting gap. Together, these findings establish pentagonal $X$Te$_2$ as a new class of two-dimensional nonsymmorphic semiconductors with symmetry-enforced Dirac band dispersions and provide fundamental insight into the interplay between crystal symmetry and local bonding during the transformation from the hexagonal parent phase.

\section{Methods}
\subsection{First-Principles Calculations}
All density-functional theory (DFT) calculations were performed using the Vienna \textit{ab initio} Simulation Package (VASP) \cite{Kresse1996} within the projector-augmented wave (PAW) formalism \cite{Blochl1994,Kresse1999} and the Perdew--Burke--Ernzerhof (PBE) generalized gradient approximation \cite{PBE1996}. We employed a plane-wave kinetic energy cutoff of 520~eV, validated by convergence tests reported in Sec.~S1 of the Supplementary Information, and sampled the Brillouin zone of the primitive unit cells with a $\Gamma$-centered $12 \times 12 \times 1$ $k$-point mesh. A 20~\AA\ vacuum layer was added along the out-of-plane direction to prevent interactions between periodic lattices. Structures were fully relaxed using a conjugate-gradient algorithm until the Hellmann--Feynman forces on each atom fell below 0.01~eV/\AA, with an electronic self-consistency criterion of $10^{-8}$~eV. We also tested an on-site Coulomb correction for NiTe$_2$, the only compound here with a partially filled $3d$ shell, which leaves the pentagonal phase nonmagnetic with the band gap and the Te--Te dimer length essentially unchanged (Supplementary Table~S1).

Dynamical stability was assessed by calculating phonon dispersions via the finite-displacement method as implemented in the Phonopy package \cite{Togo2015}. Interatomic force constants were computed in VASP using the base parameters described above. To ensure the proper convergence of acoustic modes, we constructed $4 \times 4 \times 1$ supercells for PdTe$_2$ and NiTe$_2$, and a larger $6 \times 6 \times 1$ supercell for PtTe$_2$. These supercell force calculations utilized a $\Gamma$-centered $2 \times 2 \times 1$ $k$-point mesh and an atomic displacement amplitude of 0.01~\AA. 
Structural robustness at finite temperatures was examined using \textit{ab initio} molecular dynamics (AIMD) simulations in the canonical ($NVT$) ensemble with a Nosé--Hoover thermostat. For each pentagonal $X\mathrm{Te}_2$ monolayer, a $3 \times 3 \times 1$ supercell containing 54 atoms was simulated at 300 and 600~K using only the $\Gamma$ point for Brillouin-zone sampling. Each trajectory was propagated for 5~ps using a time step of 2~fs. Spin--orbit coupling was not included in the AIMD calculations.
\subsection{Structural Relaxation and Transformation Pathways}
\label{sec:interp}
To trace the structural transition from the hexagonal to the pentagonal phase, we constructed a continuous pathway connecting the two fully relaxed endpoint structures, parametrized by the interpolation parameter $\lambda$, where $\lambda = 0$ corresponds to the hexagonal phase and $\lambda = 1$ to the pentagonal phase. Because the primitive hexagonal unit cell contains three atoms (PdTe$_2$), a $2 \times 1$ rectangular supercell with six atoms (Pd$_2$Te$_4$) was constructed to match the stoichiometry and atom count of the pentagonal unit cell, enabling a one-to-one atomic mapping.
Along this pathway, the in-plane lattice vectors were interpolated linearly according to
\begin{equation}
\mathbf{a}(\lambda) = (1-\lambda)\,\mathbf{a}^{\mathrm{hex}} + \lambda\,\mathbf{a}^{\mathrm{penta}},
\end{equation}
where $\mathbf{a}$ represents the in-plane lattice constant, and the fractional atomic coordinates were interpolated as
\begin{equation}
\mathbf{r}_i(\lambda) = \mathbf{r}_i^{\mathrm{hex}} + \lambda\,\Delta_i,
\end{equation}
where $\Delta_i$ represents the displacement vector between the two endpoint structures. To ensure that atoms follow the shortest displacement under periodic boundary conditions, each component of $\Delta_i$ was mapped into the interval $[-0.5,\,0.5]$ prior to interpolation. 
The correspondence between the rectangular supercell and the pentagonal cell is not unique, since the two inequivalent orientations of the supercell relative to the pentagonal lattice generate distinct pathways between the same pair of endpoint structures. We constructed both and adopted the one with the lower energy profile throughout, which is the correspondence in which the long axis of the supercell maps onto the long axis of the pentagonal cell. 
At each value of $\lambda$, the lattice parameters were fixed at their interpolated values while the internal atomic coordinates were relaxed to minimize the residual forces. Electronic band structures were subsequently computed for the relaxed intermediate configurations to track the evolution of the electronic structure and band gap along the transformation pathway.

In addition to this interpolation pathway, the minimum-energy path connecting the same two endpoint structures was obtained with the climbing-image solid-state
nudged elastic band method \cite{Henkelman2000,Sheppard2012} as implemented in
the VTST extension to VASP, in which the atomic positions and the cell degrees of
freedom are optimized simultaneously. The band comprised eleven images numbered 0
to 10, with the two fully relaxed endpoints held fixed and the nine interior
images initialized from the linear interpolation. Spin--orbit coupling was
included throughout, and the plane-wave cutoff, PAW potentials,
exchange--correlation functional, $k$-point mesh and electronic convergence
criterion were identical to those used for the endpoint relaxations. Full computational details are given in Sec.~S4 of the Supplementary Information.

\section{Results and Discussion}
\subsection{Structural Stability of Pentagonal \textit{X}Te$_2$}
\label{sec:stability}

\begin{figure*}[htbp]
    \centering
    \includegraphics[width=\textwidth]{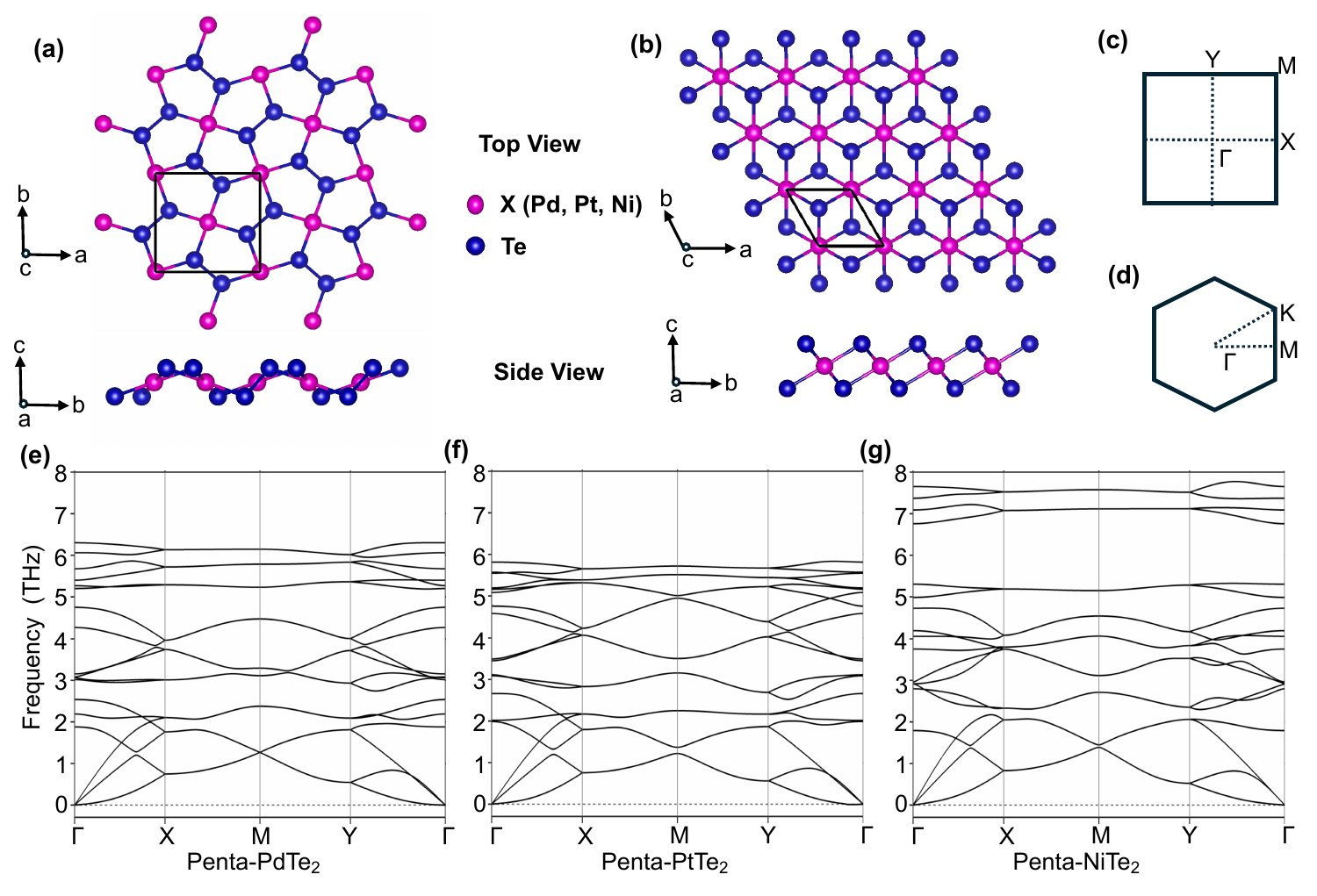}
    \caption{Crystal structures and Brillouin zones of pentagonal and hexagonal $X$Te$_2$ ($X = \mathrm{Pd}, \mathrm{Pt}, \mathrm{Ni}$) monolayers, and phonon dispersions of the pentagonal phase.
(a,b) Top and side views of the pentagonal (a) and hexagonal (b) structures. Magenta and blue spheres denote X and Te atoms, respectively.
(c,d) Rectangular Brillouin zone of the pentagonal phase (c) and hexagonal Brillouin zone of the 1T phase (d).
(e--g) Phonon dispersions of pentagonal PdTe$_2$, PtTe$_2$, and NiTe$_2$, respectively.}

\label{fig:structure}

\end{figure*}

The optimized crystal structures of the pentagonal and hexagonal (1T) polymorphs of $X$Te$_2$ ($X = \mathrm{Pd}, \mathrm{Pt}, \mathrm{Ni}$) are shown in Figs.~\ref{fig:structure}(a) and \ref{fig:structure}(b). The hexagonal phase crystallizes in the $P\bar{3}m1$ space group with three atoms per primitive unit cell ($X$Te$_2$), where each $X$ atom sits at the center of an octahedron formed by six Te atoms. These edge-sharing $X$Te$_6$ octahedra form a symmetric Te--$X$--Te trilayer without in-plane dimerization, as seen in the side view of Fig.~\ref{fig:structure}(b). The corresponding Brillouin zone is hexagonal, with high-symmetry points $\Gamma$, M, and K (Fig.~\ref{fig:structure}(d)).

The pentagonal phase presents a distinct structural motif. It belongs to the layer group $p2_1/b11$ (3D space group $P2_1/c$) and contains six atoms per unit cell ($X_2$Te$_4$). The Bravais lattice symmetry is reduced from hexagonal to rectangular, yielding inequivalent in-plane lattice constants and a rectangular Brillouin zone with high-symmetry points $\Gamma$, X, M, and Y (Fig.~\ref{fig:structure}(c)). Unlike the flat 1T structure, the pentagonal phase adopts a puckered geometry characterized by Te--Te dimers that bridge adjacent pentagonal rings formed by three Te and two $X$ atoms. This dimerization lowers the $X$ coordination from sixfold to fourfold and modifies the local bonding environment. The optimized lattice parameters and electronic band gaps for both polymorphs are summarized in Table~\ref{tab:lattice_gap}. The hexagonal phases have lattice constants in the range $3.77$--$4.02$~\AA, while the pentagonal phases adopt nearly square unit cells with $a \approx 6.26$--$6.46$~\AA\ and $b \approx 5.95$--$6.16$~\AA. 
For comparison, epitaxial pentagonal PdTe$_2$ grown on Pd(100) adopts the square $(\sqrt{5}\times\sqrt{5})R26.6^{\circ}$ surface cell of the substrate, 6.15~\AA\ on a side \cite{Liu2024}, which relative to our free-standing values is a compression of 4.80\% along $a$ and of only 0.16\% along $b$. Constraining the monolayer to that cell leaves the $P2_1/c$ space group and the symmetry-enforced Dirac points intact and reduces the band gap to 1.11~eV (Supplementary Table~S5 and Fig.~S10).

\begin{table}[b]
\caption{Optimized in-plane lattice constants and electronic band gaps ($E_g$) 
of monolayer PdTe$_2$, PtTe$_2$, and NiTe$_2$ in the 1T (hexagonal) and pentagonal phases. 
The hexagonal phase has $a = b$ by symmetry. 
All structures are fully relaxed and band gaps are computed including SOC.}
\label{tab:lattice_gap}
\begin{ruledtabular}
\begin{tabular}{lcccc}
Material & Phase & $a$ (\AA) & $b$ (\AA) & $E_g$ (eV) \\
\hline
\multirow{2}{*}{PdTe$_2$} 
    & 1T    & 4.02 & 4.02 & 0 \\
    & Penta & 6.46 & 6.16 & 1.23 \\
\hline
\multirow{2}{*}{PtTe$_2$} 
    & 1T    & 4.02& 4.02& 0.37 \\
    & Penta & 6.41 & 6.12 & 1.32 \\
\hline
\multirow{2}{*}{NiTe$_2$} 
    & 1T    & 3.77& 3.77& 0 \\
    & Penta & 6.26 & 5.95 & 0.92 \\
\end{tabular}
\end{ruledtabular}
\end{table}

To verify dynamical stability, we computed phonon dispersions for all three compounds in the pentagonal phase (Figs.~\ref{fig:structure}(e)--\ref{fig:structure}(g)). No imaginary phonon frequencies appear throughout the Brillouin zone, confirming that pentagonal PdTe$_2$, PtTe$_2$, and NiTe$_2$ correspond to dynamically stable local energy minima. Although the hexagonal phases remain energetically favored, the pentagonal polymorphs are robust metastable phases, providing well-defined structural platforms for the symmetry-driven electronic phenomena. To complement the zero-temperature phonon analysis, the structural robustness of the pentagonal phases at finite temperatures was evaluated using AIMD simulations at 300 and 600~K. As shown in Supplementary Figs.~S1--S3, once each system has equilibrated at the
set temperature the potential-energy trajectories of all three compounds exhibit
bounded thermal fluctuations without systematic drift over the 5~ps simulations.
To characterize the dimer network throughout each trajectory, we measured the eighteen Te--Te dimer separations of the
supercell in every configuration recorded after equilibration (Supplementary
Table~S2). At 300~K the mean separations of 2.81, 2.83 and 2.84~\AA\ for
PdTe$_2$, PtTe$_2$ and NiTe$_2$ reproduce the relaxed values, and at 600~K they
increase only to 2.82, 2.84 and 2.87~\AA. The dimerized bonding motif that
underlies the semiconducting gap is therefore retained, and these results support
the structural robustness of the pentagonal phases at 300 and 600~K over the
simulated timescale.

\subsection{Electronic Structures of Hexagonal and Pentagonal Phases}

\begin{figure*}[htbp]
    \centering
    \includegraphics[width=\textwidth]{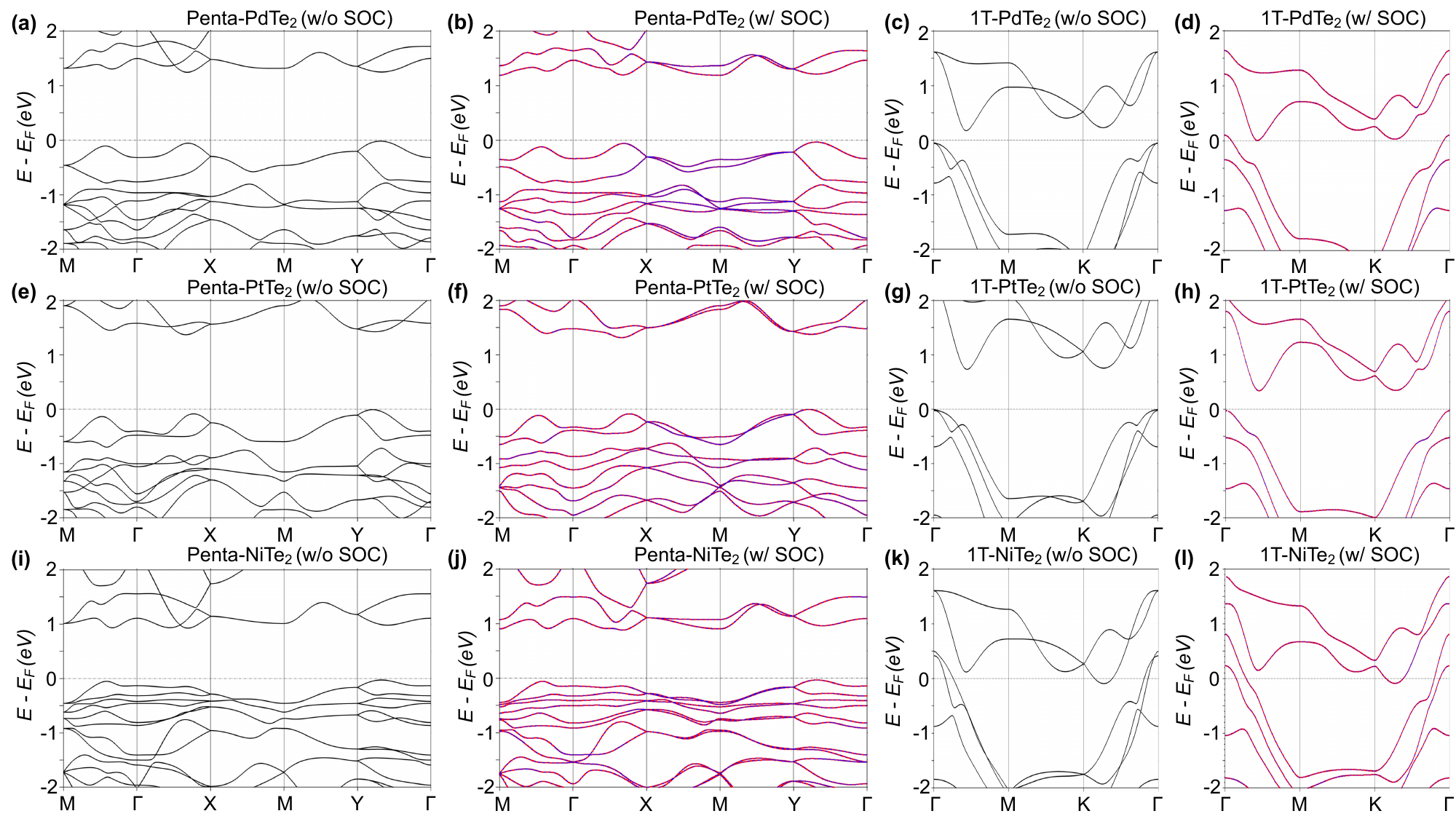} 
    \caption{Electronic band structures of pentagonal (Penta) and hexagonal (1T) $X$Te$_2$ 
    ($X = \mathrm{Pd}, \mathrm{Pt}, \mathrm{Ni}$) monolayers with and without spin--orbit coupling (SOC).
    (a,b) Penta-PdTe$_2$ without and with SOC, respectively; 
    (c,d) 1T-PdTe$_2$ without and with SOC; 
    (e,f) Penta-PtTe$_2$ without and with SOC; 
    (g,h) 1T-PtTe$_2$ without and with SOC; 
    (i,j) Penta-NiTe$_2$ without and with SOC; 
    (k,l) 1T-NiTe$_2$ without and with SOC.
    Energies are referenced to the Fermi level.
    Black lines denote calculations without SOC. For the SOC calculations, red and blue lines indicate the twofold Kramers-degenerate bands.}
    \label{fig:bands}
\end{figure*}

The electronic band structures of all three $X$Te$_2$ compounds are shown in Fig.~\ref{fig:bands} for both the hexagonal and pentagonal phases, computed with and without spin--orbit coupling (SOC). A clear contrast emerges between the two polymorphs. In the hexagonal 1T phase, PdTe$_2$ and NiTe$_2$ exhibit semimetallic behavior, with valence and conduction bands overlapping near the Fermi level (Figs.~\ref{fig:bands}(c,d) and \ref{fig:bands}(k,l)). PtTe$_2$ differs slightly by exhibiting a small indirect band gap of 0.37~eV (Fig.~\ref{fig:bands}(h)), placing it in the narrow-gap semiconducting regime. In all three hexagonal systems, the inclusion of SOC lifts certain orbital degeneracies and modifies the band dispersion, but does not produce symmetry-protected boundary degeneracies or Dirac crossings. Despite minor quantitative differences in band dispersion and gap size, the hexagonal phases share a similar electronic structure near the Fermi level and lack the nonsymmorphic band features observed in the pentagonal phase.

The pentagonal polymorphs display fundamentally different electronic structures from their hexagonal counterparts. Without SOC, the band structures of pentagonal PdTe$_2$, PtTe$_2$, and NiTe$_2$ exhibit band degeneracies along the entire Brillouin-zone boundary paths X--M and M--Y (Figs.~\ref{fig:bands}(a,e,i)). These degeneracies form nodal lines pinned to the zone boundary (Sec.~\ref{sec:dirac}). When SOC is included, the degeneracy is lifted along the boundary segments but persists at the high-symmetry points X and Y, where fourfold-degenerate Dirac crossings appear (Figs.~\ref{fig:bands}(b,f,j)). The symmetry origin of these Dirac points is analyzed in detail in Sec.~\ref{sec:dirac}. In addition to these zone-boundary band features, the pentagonal phases are semiconducting with substantially larger band gaps of 1.23~eV (PdTe$_2$), 1.32~eV (PtTe$_2$), and 0.92~eV (NiTe$_2$). The emergence of a robust $\sim 1$~eV gap across the family reflects the reduced coordination and Te--Te dimerization introduced in the previous subsection, which modifies the $d$--$p$ hybridization and drives bonding--antibonding splitting. The microscopic origin of this gap opening is quantified through the structural interpolation analysis in Sec.~\ref{sec:transition}.

\subsection{Symmetry-Enforced Dirac Fermions in Pentagonal \textit{X}Te$_2$}
\label{sec:dirac}

\begin{figure*}[htbp]
    \centering
    \includegraphics[width=\textwidth]{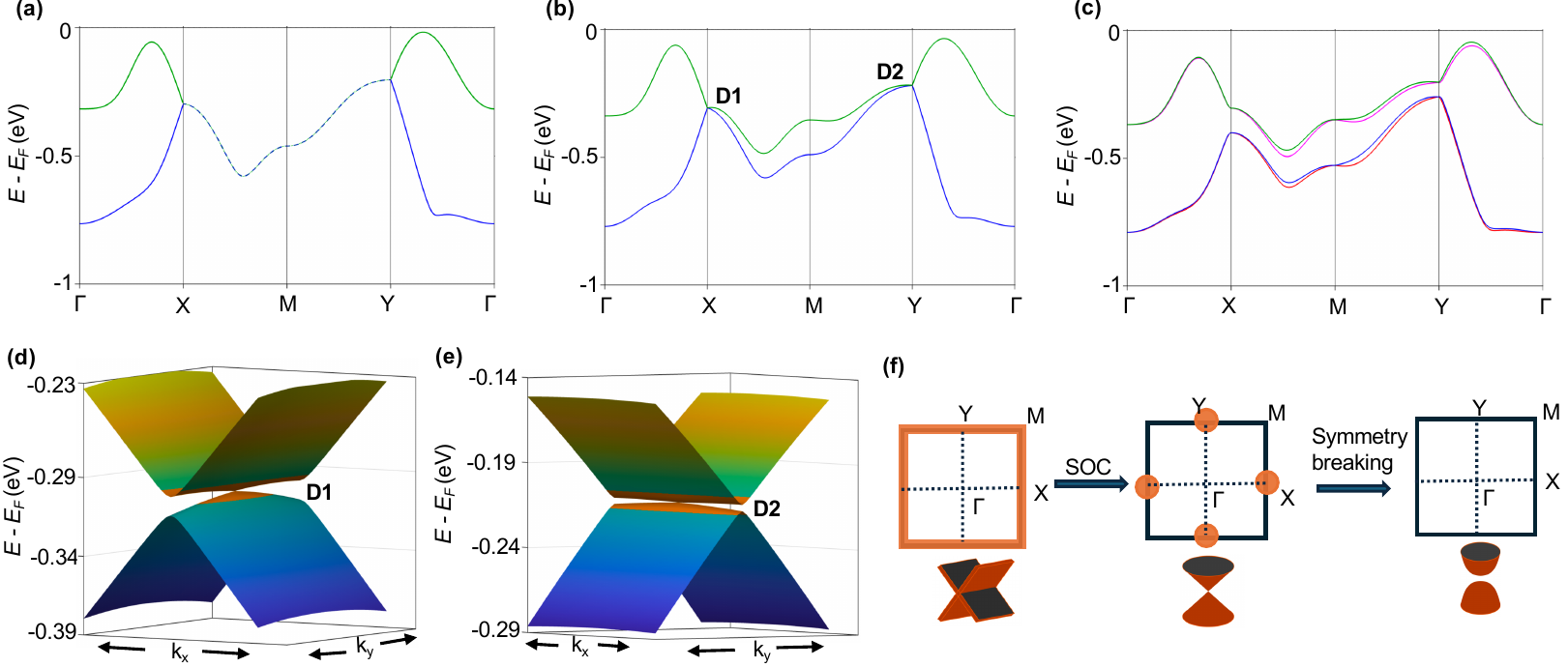} 
    \caption{Effects of spin--orbit coupling (SOC) and symmetry breaking on the electronic structure of pentagonal PdTe$_2$.
(a) Electronic band structure without SOC.
 The two highest occupied bands are
highlighted in green and blue and become degenerate along the X--M and M--Y zone-boundary paths, forming the fourfold nodal lines discussed in Sec.~\ref{sec:dirac}. Alternating dashed segments in the two
colors indicate the overlapping degenerate bands.
(b) Electronic band structure with SOC. The corresponding
Kramers-degenerate branches are separated at generic points along the zone boundary but meet at the fourfold-degenerate Dirac points D1 and D2 at X and Y, respectively, as enforced by the nonsymmorphic
symmetry.
(c) Electronic band structure of the symmetry-broken phase, in which atomic displacements of approximately $1.5\%$ of the in-plane lattice constant (see Supplementary Table~S3) lift the Dirac degeneracy and gap the Dirac crossings. The four colors distinguish the resulting band branches.
(d, e) Three-dimensional DFT band energy surfaces calculated in the vicinity of the Dirac points labeled D1 at X (d) and D2 at Y (e), corresponding to the crossings highlighted in panel (b).
(f) Schematic evolution of the band configuration from the case without SOC to the SOC-included and symmetry-broken phase.
Energies are referenced to the Fermi level.}
    \label{fig:topology}
\end{figure*}

As shown in the preceding subsection, robust band crossings occur at the time-reversal invariant momenta $\mathrm{X}=(\pi,0)$ and $\mathrm{Y}=(0,\pi)$, with momenta quoted in dimensionless form $(k_xa,k_yb)$. We now demonstrate that these crossings are not accidental but are strictly enforced by the nonsymmorphic symmetry of the pentagonal monolayer (space group $P2_1/c$, No.~14; layer group $p2_1/b11$, No.~17). The monolayer is nonmagnetic and centrosymmetric; therefore, both time-reversal symmetry ($\hat{T}$) and inversion symmetry ($\hat{P}$) are preserved. In the absence of SOC, the band structure exhibits nodal lines along the X--M and M--Y zone boundaries, as shown in Fig.~\ref{fig:topology}(a). When SOC is included, these boundary degeneracies are lifted along the zone edges, yet fourfold-degenerate Dirac crossings persist at X and Y (Fig.~\ref{fig:topology}(b)). The nonsymmorphic symmetry enforces the same degeneracy for every band at X and Y, in both the valence and conduction manifolds, as is visible in Figs.~\ref{fig:bands}(b,f,j). The crossings examined in Figs.~\ref{fig:topology}(d,e) are the bands closest to the Fermi level. Three-dimensional band-energy surfaces near these points  reveal conical band crossings centered at X and Y, confirming the Dirac nature of these nodes. When the protecting symmetries (inversion and glide) are explicitly broken by lattice distortions, the Dirac crossings gap out, and the fourfold degeneracy splits into spin-split bands (Fig.~\ref{fig:topology}(c)), confirming the
symmetry-protection mechanism of band degeneracies. This symmetry hierarchy is summarized schematically in Fig.~\ref{fig:topology}(f). To quantify the dispersion near the symmetry-enforced Dirac nodes, we extract the Dirac velocities for the band crossings at X and Y. For pentagonal PdTe$_2$, the Dirac cone at X is strongly anisotropic, with $v_D^{\Gamma \rightarrow X} \approx 3.8 \times 10^5$~m/s and $v_D^{X \rightarrow M} \approx 3.0 \times 10^4$~m/s, yielding an anisotropy ratio exceeding one order of magnitude. An even larger anisotropy is found at Y, where $v_D^{Y \rightarrow \Gamma} \approx 3.4 \times 10^5$~m/s while $v_D^{M \rightarrow Y} \approx 9.4 \times 10^3$~m/s. In both cases, the high-velocity direction points toward $\Gamma$, reflecting the highly anisotropic character of the Dirac cones. Comparable anisotropy is obtained for PtTe$_2$ and NiTe$_2$. A quantitative comparison of the extracted Dirac velocities for all three compounds is given in Supplementary Fig.~S4.

The symmetry argument of band degeneracies proceeds as follows. The monolayer penta-$X$Te$_2$ preserves inversion $\hat P$ and a nonsymmorphic glide symmetry $\hat G$:
\begin{equation}
\hat P:(x,y,z)\rightarrow(-x,-y,-z),
\end{equation}
\begin{equation}
\hat G=\{M_y \mid \tfrac{1}{2},\tfrac{1}{2},0\}:(x,y,z)\rightarrow
\left(x+\tfrac{1}{2},-y+\tfrac{1}{2},z\right),
\end{equation}
where the mirror reflection $M_y$ is accompanied by a fractional translation, making $\hat G$ a nonsymmorphic operation.

The key observation is that $\hat P$ and $\hat G$ do not commute: applying them in opposite orders yields results that differ by a lattice translation. Acting on real-space coordinates,
\begin{align}
(x,y,z)\xrightarrow{\,\hat G\,}\left(x+\tfrac{1}{2},-y+\tfrac{1}{2},z\right)
\xrightarrow{\,\hat P\,}\left(-x-\tfrac{1}{2},\,y-\tfrac{1}{2},\,-z\right), \label{eq:PG_path}\\
(x,y,z)\xrightarrow{\,\hat P\,}(-x,-y,-z)
\xrightarrow{\,\hat G\,}\left(-x+\tfrac{1}{2},\,y+\tfrac{1}{2},\,-z\right). \label{eq:GP_path}
\end{align}
Comparing Eqs.~\eqref{eq:PG_path} and~\eqref{eq:GP_path}, the two final coordinates are related by a primitive translation by one unit cell along both in-plane directions. Denoting this translation by $\hat t_{(1,1,0)}$, we obtain the operator identity
\begin{equation}
\hat G\hat P = \hat t_{(1,1,0)}\,\hat P\hat G.
\end{equation}
For Bloch states, $\hat t_{(1,1,0)}$ contributes a momentum-dependent phase, so the above becomes
\begin{equation}
\hat G\hat P = e^{-i(k_x+k_y)}\,\hat P\hat G.
\end{equation}

On the rectangular Brillouin-zone boundary,
\begin{align}
\mathrm{X}=(\pi,0),\quad \mathrm{Y}=(0,\pi):\qquad & e^{-i(k_x+k_y)}=-1 \;\Rightarrow\; \{\hat P,\hat G\}=0, \\
\mathrm{M}=(\pi,\pi):\qquad & e^{-i(k_x+k_y)}=+1 \;\Rightarrow\; [\hat P,\hat G]=0.
\end{align}

Let $|\psi\rangle$ be a Bloch eigenstate at X or Y satisfying
\begin{equation}
\hat G|\psi\rangle = g|\psi\rangle.
\end{equation}
Using $\{\hat P,\hat G\}=0$ at X and Y,
\begin{equation}
\hat G(\hat P|\psi\rangle) = -\hat P(\hat G|\psi\rangle) = -g\,\hat P|\psi\rangle .
\end{equation}
Since X and Y are time-reversal invariant momenta, inversion maps $\bm{k}$ onto itself modulo a reciprocal lattice vector. Therefore $|\psi\rangle$ and $\hat P|\psi\rangle$ occur at the same crystal momentum and carry opposite glide eigenvalues. This enforces a twofold degeneracy at X and Y even in the spinless orbital limit.

With SOC, the antiunitary symmetry $\hat P\hat T$ satisfies $(\hat P\hat T)^2=-1$, enforcing a twofold Kramers degeneracy throughout the Brillouin zone. At X and Y, this $\hat P\hat T$-protected doublet combines with the glide--inversion enforced doubling, giving $2\times 2 = 4$, hence a symmetry-enforced fourfold Dirac degeneracy at X and Y.

Beyond the high-symmetry points, the nonsymmorphic symmetry also protects degeneracies at the Brillouin zone boundary in the spinless orbital limit. In this limit,
\(\hat T^2=+1\). Squaring the glide gives
\begin{equation}
\hat G^2=\hat t_{(1,0,0)}
\quad\Rightarrow\quad
\hat G^2=e^{-ik_x}.
\end{equation}
Along the zone edge X--M, where \(k_x=\pi\), the antiunitary operation
\(\hat{\Theta}_x\equiv \hat T\hat G\) leaves \textbf{\textit{k}} invariant and satisfies
\begin{equation}
\hat{\Theta}_x^2=(\hat T\hat G)^2=e^{-ik_x}=-1 .
\end{equation}
Thus the spinless orbital Hamiltonian has a twofold degeneracy along
X--M.

For the other zone edge, define the screw symmetry
\begin{equation}
\hat S\equiv \hat P\hat G .
\end{equation}
This operation satisfies
\begin{equation}
\hat S^2=\hat t_{(0,-1,0)}
\quad\Rightarrow\quad
\hat S^2=e^{ik_y}.
\end{equation}
Along M--Y, where \(k_y=\pi\), the antiunitary operation
\(\hat{\Theta}_y\equiv \hat T\hat S\) satisfies
\begin{equation}
\hat{\Theta}_y^2=(\hat T\hat S)^2=e^{ik_y}=-1 .
\end{equation}
Therefore the spinless orbital Hamiltonian has a symmetry-enforced
twofold nodal line along the Brillouin-zone boundary.

The physical band structure without SOC consists of two identical spin copies
of this spinless orbital Hamiltonian,
\begin{equation}
H_{\rm noSOC}(\bm{k})=H_{\rm orb}(\bm{k})\otimes\sigma_0 ,
\end{equation}
where $\sigma_0$ is the 2$\times$2 identity matrix in the spin space. Thus, when the spin degree of freedom is counted, the no-SOC nodal line is fourfold degenerate.

When SOC is included, the SOC-related terms in the Hamiltonian couple the two spin sectors and split the fourfold no-SOC
nodal line into two Kramers-degenerate branches. The remaining
twofold degeneracy is protected by \(\hat P\hat T\), since
\begin{equation}
(\hat P\hat T)^2=-1 .
\end{equation}

This can be seen explicitly from the relation between the antiunitary
operators on the zone boundary. Along X--M, the screw operation $\hat S=\hat P\hat G$ maps $(k_x,k_y)\to(-k_x,k_y)$ and hence leaves every point with $k_x=\pi$ invariant, making it a unitary
little-group operation. The antiunitary operation \(\hat T\hat G\) can be written as
\begin{equation}
\hat T\hat G=(\hat P\hat T)(\hat P\hat G)=(\hat P\hat T)\hat S .
\end{equation}
For an eigenstate of \(\hat S\),
\begin{equation}
\hat S|\psi\rangle=s|\psi\rangle ,
\end{equation}
we obtain
\begin{equation}
\hat T\hat G|\psi\rangle=(\hat P\hat T)\hat S|\psi\rangle=s^*\hat P\hat T|\psi\rangle .
\end{equation}
Thus \(\hat T\hat G|\psi\rangle\) is not an independent partner beyond the
ordinary \(\hat P\hat T|\psi\rangle\) Kramers partner. Hence SOC leaves only a
twofold Kramers degeneracy at a generic point on X--M, rather than
the fourfold no-SOC nodal-line degeneracy.

Similarly, along M--Y, the glide $\hat G$ maps $(k_x,k_y)\to(k_x,-k_y)$ and
leaves every point with $k_y=\pi$ invariant, so it is the relevant unitary
little-group operation there.
Since \(\hat S=\hat P\hat G\), one has
\begin{equation}
\hat T\hat S=\hat T(\hat P\hat G)=(\hat P\hat T)\hat G .
\end{equation}
For a glide eigenstate
\begin{equation}
\hat G|\psi\rangle=g|\psi\rangle ,
\end{equation}
one obtains
\begin{equation}
\hat T\hat S|\psi\rangle=(\hat P\hat T)\hat G|\psi\rangle=g^*\hat P\hat T|\psi\rangle .
\end{equation}
Therefore \(\hat T\hat S|\psi\rangle\) is also not an independent partner beyond
\(\hat P\hat T|\psi\rangle\). SOC therefore lifts the fourfold no-SOC boundary nodal
line along generic points of X--M and M--Y, leaving twofold
Kramers-degenerate branches.

At the time-reversal invariant momenta (TRIM) X and Y, the situation is different. There,
both inversion and glide belong to the little group and satisfy
\begin{equation}
\{\hat P,\hat G\}=0 .
\end{equation}
Thus \(\hat P|\psi\rangle\) carries the opposite glide eigenvalue from
\(|\psi\rangle\). With SOC, each of these two glide-related states also has
a \(\hat P\hat T\) Kramers partner. Consequently,
\begin{equation}
2_{\rm glide/inversion}\times 2_{\rm Kramers}=4 ,
\end{equation}
so the fourfold Dirac points at X and Y remain symmetry protected.

This argument relies only on the nonsymmorphic layer group $p2_1/b11$, inversion symmetry, and time-reversal symmetry, and is therefore independent of the chemical identity of the transition metal. Since pentagonal PtTe$_2$ and NiTe$_2$ share the same space group and are nonmagnetic, the same symmetry constraints apply: fourfold Dirac points are enforced at X and Y in all three compounds, as confirmed by the band structures in Figs.~\ref{fig:bands}(f) and~\ref{fig:bands}(j). The pentagonal $X$Te$_2$ family thus constitutes a class of two-dimensional semiconductors with nonsymmorphic Dirac crossings.

\subsection{Semimetal-to-Semiconductor Transition: Energetics and the Microscopic Role of Te--Te Dimerization}
\label{sec:transition}

The preceding sections established that the hexagonal and pentagonal polymorphs of $X$Te$_2$ possess fundamentally different electronic characters: the former is semimetallic, whereas the latter is semiconducting with symmetry-enforced Dirac nodes. To understand how one evolves into the other, we trace the structural
transformation pathway defined in Sec.~\ref{sec:interp}, where the
interpolation parameter $\lambda$ connects the relaxed hexagonal
$2\times1$ supercell ($\lambda=0$) to the relaxed pentagonal structure
($\lambda=1$). We focus on PdTe$_2$, the experimentally realized member
of the family~\cite{Liu2024}, as a representative case; since PtTe$_2$
and NiTe$_2$ undergo the identical symmetry reduction
($P\bar{3}m1 \rightarrow P2_1/c$) and develop the same Te--Te dimer
motif, qualitatively similar transformation behavior is expected across
the family.

\begin{figure*}[htbp]
    \centering
    \includegraphics[width=\textwidth]{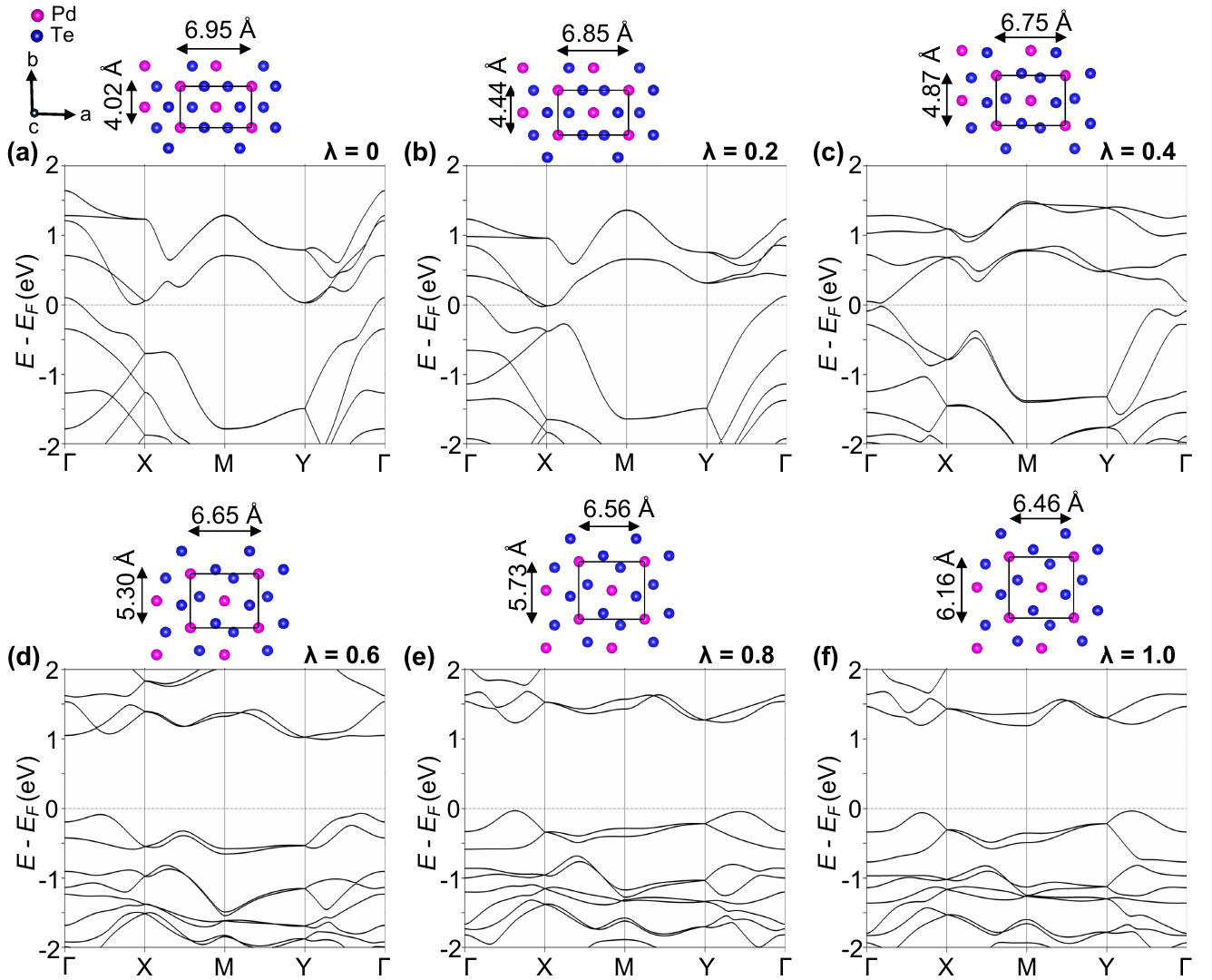} 
    \caption{Evolution of the crystal and electronic structures of PdTe$_2$ along
the hexagonal-to-pentagonal transformation path. (a)--(f) For each value of the
interpolation parameter $\lambda$, where $\lambda = 0$ corresponds to the
hexagonal phase and $\lambda = 1$ to the pentagonal phase, the relaxed crystal
structure (upper panel) and the corresponding electronic band structure
calculated with spin--orbit coupling (SOC) (lower panel) are shown. The
in-plane lattice constants $a$ and $b$ of the rectangular supercell are
indicated in each structural panel. Magenta and blue spheres denote Pd and Te
atoms, respectively. The system evolves continuously from a semimetallic state
at $\lambda = 0$ to a semiconducting state at $\lambda = 1$, with a band gap opening for $\lambda \gtrsim 0.4$. Energies are referenced to the Fermi
level.}
    \label{fig:transition}
\end{figure*}

Figure~\ref{fig:transition} presents the relaxed crystal structures and
corresponding electronic band structures (including SOC) of PdTe$_2$ at
representative values of $\lambda$. At $\lambda=0$, the system adopts a flat trilayer geometry expressed in a rectangular supercell but retains the underlying atomic arrangement of the 1T hexagonal phase, exhibiting its characteristic semimetallic behavior. As $\lambda$ increases, the lattice parameters evolve toward a more square-like aspect ratio while the atomic sublayers develop increasing out-of-plane puckering. Throughout the range $0 \le \lambda \lesssim 0.4$, the threefold rotational symmetry characteristic of the parent $P\bar{3}m1$ structure is progressively broken, leading to the splitting of band degeneracies inherited from the hexagonal structure. The system nevertheless remains semimetallic throughout this regime.

\begin{figure*}[htbp]
\centering
\includegraphics[width=\textwidth]{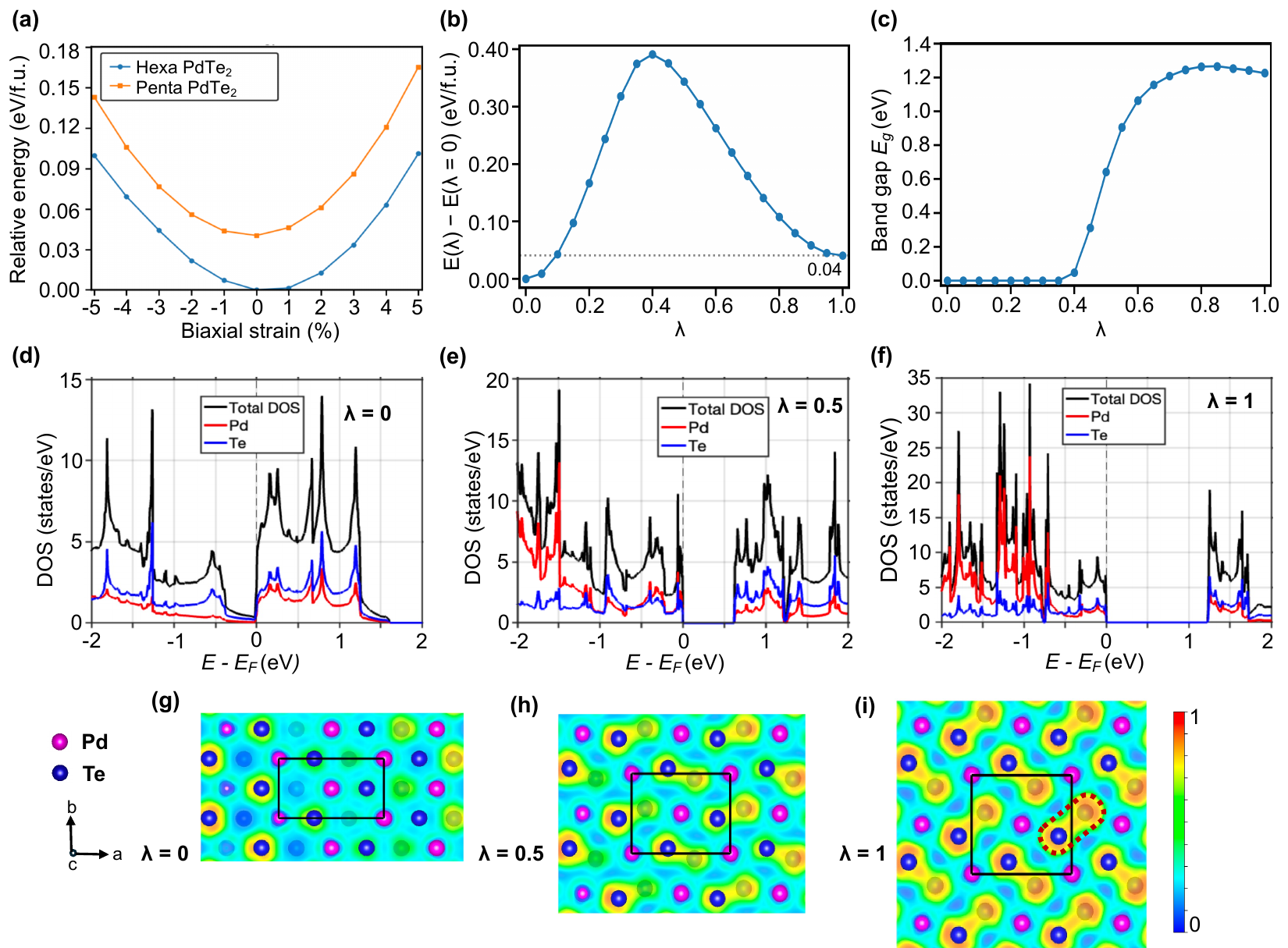}
\caption{Energetics, band-gap evolution, density of states, and electron localization function (ELF) of PdTe$_2$ along the hexagonal-to-pentagonal transformation pathway. (a) Relative total energy per formula unit (f.u.), referenced to the equilibrium hexagonal phase, for hexagonal and pentagonal PdTe$_2$ under biaxial strain. No energy crossing is observed within the considered strain range.
(b) Energy profile $\Delta E(\lambda)$ along the structural interpolation path connecting the hexagonal ($\lambda=0$) and pentagonal ($\lambda=1$) phases. (c) Evolution of the electronic band gap $E_g$ as a function of the
interpolation parameter $\lambda$. A band gap opens for $\lambda \gtrsim 0.4$. (d)--(f) Density of states (DOS) for PdTe$_2$ at (d) $\lambda = 0$, (e) $\lambda = 0.5$, and (f) $\lambda = 1$. The total DOS is shown in black, while Pd $d$ and Te $p$ contributions are shown in red and blue, respectively. With increasing $\lambda$, Pd $d$ states shift toward deeper valence energies (below $\sim -0.65$~eV), while Te $p$ states remain dominant near the band edges. Note the different vertical scales in panels (d) to (f). (g)--(i) Top views of the electron localization function (ELF) for PdTe$_2$ at (g) $\lambda = 0$, (h) $\lambda = 0.5$, and (i) $\lambda = 1$. The color mapping indicates the degree of electron localization, where values approaching 1 (red) represent strong electron localization and covalent bonding, while lower values (blue) indicate weak localization. As $\lambda$ increases, the ELF progressively localizes between neighboring Te atoms, providing direct evidence for the development of Te--Te covalent bonding. The dashed red outline in panel (i) highlights one of the fully formed Te--Te dimers in the pentagonal phase.}
\label{fig:energetics}
\end{figure*}

A distinct electronic transition occurs near $\lambda \approx 0.4$, where a finite band gap begins to open (Fig.~\ref{fig:energetics}(c)).
Beyond this critical distortion the gap rises steeply and saturates above $\lambda \approx 0.8$, reaching the fully developed semiconducting state of the pentagonal phase with $E_g = 1.23$~eV. The transition therefore does not coincide with the symmetry reduction, which already occurs for any $\lambda > 0$, but emerges only at an intermediate structural threshold.

To identify the driving force behind this threshold behavior, we first examine whether the hexagonal-to-pentagonal transition can be induced by a simple external lattice distortion. Figure~\ref{fig:energetics}(a)  compares the total energies of the hexagonal and pentagonal phases under biaxial strain ranging from $-5\%$ to $+5\%$. The two energy curves do not intersect within this range, indicating that the pentagonal phase cannot be stabilized by a simple lattice strain and instead requires internal structural reorganization.
The energy landscape along the $\lambda$ pathway (Fig.~\ref{fig:energetics}(b))
reveals a finite barrier separating the two endpoints, rising from the hexagonal
phase to a maximum of 0.391~eV per formula unit at $\lambda = 0.40$ and decreasing
toward the pentagonal structure, which lies 0.04~eV per formula unit above the
hexagonal phase in agreement with the zero-strain separation in
Fig.~\ref{fig:energetics}(a). Because the cell parameters are held at their
interpolated values at each $\lambda$, this pathway is not the minimum-energy
path, so its maximum is an upper bound on the saddle-point energy. We therefore
also performed a climbing-image solid-state nudged elastic band calculation, in
which the atomic positions and the cell relax simultaneously. It gives a saddle
point of 0.357~eV per formula unit at image 4 of the eleven-image band, 8.8\% below the maximum of the interpolation profile, and lies below that profile at every intermediate image (Supplementary Figs.~S6 and S7). This is the activation barrier for a concerted transformation of the periodic cell. The pentagonal phase
therefore occupies a metastable local minimum separated from the hexagonal phase
by a barrier of order 0.3~eV per formula unit, consistent with the dynamical and
thermal stability established in Sec.~\ref{sec:stability}.

\begin{table}[b]
\caption{Evolution of the minimum Te--Te distance and average Pd--Te distance as a function of the structural interpolation parameter $\lambda$ along the hexagonal-to-pentagonal transformation pathway.}
\label{tab:bond_lengths}
\begin{ruledtabular}
\begin{tabular}{lcc}
$\lambda$ & $d_{\mathrm{Te-Te}}$ (min) (\AA) & $d_{\mathrm{Pd-Te}}$ (avg) (\AA) \\
\hline
0.0 & 3.62 & 2.70 \\
0.3 & 3.13 & 2.75 \\
0.4 & 2.97 & 2.69 \\
0.5 & 2.89 & 2.66 \\
0.6 & 2.86 & 2.65 \\
0.7 & 2.83 & 2.64 \\
0.8 & 2.82 & 2.63 \\
1.0 & 2.81 & 2.63 \\
\end{tabular}
\end{ruledtabular}
\end{table}

To identify the structural trigger for the gap opening, we examine the evolution of interatomic distances along the transformation path (Table~\ref{tab:bond_lengths}). The average Pd--Te bond length changes only modestly across the full range of $\lambda$. In contrast, the minimum Te--Te distance contracts sharply from 3.62~\AA\ at $\lambda=0$ to 2.97~\AA\
at $\lambda=0.4$, which coincides with the opening of the band gap (Fig.~\ref{fig:energetics}(c)).
The separation continues to decrease and falls below the intrachain Te--Te bond length of elemental trigonal Te, 2.835~\AA~\cite{Adenis1989}, near $\lambda \approx 0.7$, at which point the gap has reached 99\% of its value in the fully relaxed pentagonal phase, and the fully formed dimer length of 2.81~\AA\ at $\lambda=1$ matches this ordinary covalent bond. This direct correlation indicates that Te--Te dimer formation acts as the structural trigger for the semimetal-to-semiconductor transition.

The electron localization function (ELF) provides direct real-space evidence for the bonding reconstruction associated with Te--Te dimerization (Figs.~\ref{fig:energetics}(g)--\ref{fig:energetics}(i)). At $\lambda=0$, the ELF shows no pronounced localization between neighboring Te atoms, consistent with the absence of Te--Te bonding in the hexagonal phase. At $\lambda=0.5$, enhanced electron localization develops between specific Te pairs, signaling the onset of bond formation. At $\lambda=1$, strong localization is concentrated in the region of the fully formed Te--Te dimers, directly confirming the formation of covalent Te--Te bonds in the pentagonal phase. This real-space evolution is consistent with dimerization-driven splitting of Te $p$ states associated with covalent Te--Te bond formation. As the Te--Te distance decreases, the Te $p$-derived states near the gap reorganize into dimer-centered states. The bonding states shift into the valence band, while the antibonding states form the conduction-band edge, thereby opening a semiconducting band gap. The corresponding VBM and CBM charge densities (Supplementary Fig.~S5) further show that the valence-band maximum exhibits bonding character between Te pairs, while the conduction-band minimum shows the antibonding counterpart of the reconstruction. Consistent with this picture, the atom-projected density of states (Figs.~\ref{fig:energetics}(d)--\ref{fig:energetics}(f)) shows substantial Te contribution near the band edges across the interpolation pathway, while Pd $d$-orbital weight is progressively redistributed into deeper valence energies (below $\sim -0.65$~eV) as dimerization strengthens. These results establish a clear microscopic picture of gap opening induced by Te--Te dimerization. The pentagonal phase therefore represents a semiconducting state that continuously evolves from the hexagonal parent structure. 
Nothing in this mechanism is specific to tellurium. The gap opens because formation of the chalcogen dimer splits the chalcogen $p$ states that straddle the Fermi level into bonding and antibonding combinations, so the same splitting is expected wherever a covalent chalcogen dimer forms in the pentagonal structure, as the semiconducting gaps of all nine pentagonal $XQ_2$ monolayers confirm (Supplementary Table~S4).

From an experimental perspective, the pentagonal polymorph is accessed through symmetry-driven epitaxy on Pd(100), as demonstrated for penta-PdTe$_2$ \cite{Liu2024}. Both Si(100) and SrTiO$_3$(100) have square surface lattices closely matched to Pd, so a Pd(100) film can be grown on either and subsequently tellurized. The predicted band features are robust to the resulting registry, since a homogeneous in-plane strain leaves every operation of the layer group invariant, so the fourfold crossings at X and Y survive strains of up to $\pm5\%$ while the gap varies between 0.65 and 1.26~eV without closing (Supplementary Table~S6). Given the sensitivity of the electronic structure to lattice metrics and dimer formation along the transformation pathway, substrate engineering offers a practical route to stabilize and tune the pentagonal phase across the $X$Te$_2$ family.

\section{Conclusion}

In summary, we establish pentagonal $X$Te$_2$ ($X=\mathrm{Pd},\mathrm{Pt},\mathrm{Ni}$) monolayers as a family of dynamically metastable semiconductors with symmetry-protected Dirac band crossings. In sharp contrast to their semimetallic or narrow-gap hexagonal 1T counterparts, these pentagonal phases exhibit sizable band gaps of 1.23~eV (PdTe$_2$), 1.32~eV (PtTe$_2$), and 0.92~eV (NiTe$_2$), extending the recently realized pentagonal PdTe$_2$ phase to a broader family of two-dimensional ditellurides. Their band topology is fundamentally dictated by rectangular nonsymmorphic symmetry (layer group $p2_1/b11$). The anticommutation of inversion and glide symmetries, combined with spin--orbit coupling, strictly enforces highly anisotropic fourfold Dirac points at the Brillouin zone boundary points X and Y. In the spinless limit, nonsymmorphic symmetry also protects nodal-line degeneracies along the zone boundaries. 
We further show that the same nonsymmorphic symmetry enforces fourfold Dirac crossings at the Brillouin-zone boundary points X and Y in the corresponding pentagonal sulfides and selenides (Supplementary Figs.~S8 and S9 and Table~S4), indicating that this symmetry protection mechanism of band crossings extends across the broader pentagonal $XQ_2$ family ($X=\mathrm{Pd},\mathrm{Pt},\mathrm{Ni}$; $Q=\mathrm{S},\mathrm{Se},\mathrm{Te}$).

The semiconducting gap of the pentagonal phase is developed during a continuous structural transformation between the hexagonal and pentagonal phases. Analysis of bond lengths, the electron localization function (ELF), VBM/CBM charge densities, and atom-projected densities of states shows that Te--Te dimerization drives a bonding--antibonding splitting of Te $p$ states, opening an energy gap of $\sim 1$~eV. Together, these results show how symmetry-enforced topology and local Te--Te bonding cooperate to define the electronic phase in pentagonal ditellurides. The coexistence of sizable semiconducting gaps, symmetry-protected Dirac crossings, and experimentally viable synthesis routes makes these materials a promising platform for exploring nonsymmorphic band topology and structural phase engineering in two dimensions. Future efforts should be directed toward the experimental growth and characterization of these pentagonal phases, as well as on probing and tuning the predicted Dirac nodes via symmetry-breaking perturbations such as substrate effects, external fields, and mechanical strain.

\section{Acknowledgements}
The work was supported by the U.S. Department of Energy, Office of Science, Office of Basic Energy Sciences, Division of Materials Science and Engineering, under Grant No. DE-SC0024294. The calculations were carried out using the high-performance computing resources maintained by Research Support Services at the University of Missouri, Columbia, MO (DOI: \href{https://doi.org/10.32469/10355/97710}{10.32469/10355/97710}).

\bibliographystyle{apsrev4-2}
\bibliography{PdTe2_paper}

\end{document}


\title{Supplementary Information \\ Symmetry-Enforced Dirac Fermions and Structural Metastability in Pentagonal Monolayers of Transition-Metal Ditellurides}

\author{Manoj Gadtoula}
\affiliation {Department of Physics and Astronomy, University of Missouri, Columbia, Missouri 65211, USA}
\author{Clayton Conner}
\affiliation {Department of Physics and Astronomy, University of Missouri, Columbia, Missouri 65211, USA}
\author{Avinash Sah}
\affiliation {Department of Physics and Astronomy, University of Missouri, Columbia, Missouri 65211, USA}
\author{Ronghao Luo}
\affiliation {Department of Physics and Astronomy, University of Missouri, Columbia, Missouri 65211, USA}
\author{Guang~Bian}
\affiliation {Department of Physics and Astronomy, University of Missouri, Columbia, Missouri 65211, USA}
\affiliation {MU Materials Science \& Engineering Institute, University of Missouri, Columbia, Missouri 65211, USA}

\maketitle

\section{Convergence tests and the on-site Coulomb correction}
\label{sec:convergence}

The 520~eV plane-wave cutoff used throughout was validated by repeating the
full relaxation and electronic-structure calculation of pentagonal NiTe$_2$,
which employs the hardest projector-augmented wave potential used in this work,
at cutoffs from 300 to 700~eV. The total energy, band gap and Te--Te dimer
length are converged well below 520~eV, and the dimer length, which is the
structural quantity central to the gap-opening mechanism, is constant to within
0.0001~\AA{} across the entire range.

Table~\ref{tab:dftu} shows the effect of an on-site Coulomb correction on
pentagonal NiTe$_2$, the only compound studied here with a partially filled
$3d$ shell. Several magnetic configurations were initialized at each value of
$U_{\mathrm{eff}}$ and all converged to the nonmagnetic solution. The Te--Te
dimer length changes by only 0.7\% over the full range and the band gap
increases monotonically, so the PBE values quoted in Table~I of the main text
are the smallest gaps the method returns.

\begin{table}[htbp]
\caption{Effect of the on-site Coulomb correction on the band gap and Te--Te
dimer length of pentagonal NiTe$_2$, using the Dudarev scheme with $J = 0$.
Every relaxed structure is nonmagnetic and retains the $P2_1/c$ space group.
These calculations are collinear and do not include spin--orbit coupling, so
the band gaps are slightly larger than the SOC-inclusive values quoted in
Table~I of the main text.}
\label{tab:dftu}
\begin{ruledtabular}
\begin{tabular}{ccc}
$U_{\mathrm{eff}}$ (eV) & $E_g$ (eV) & $d_{\mathrm{Te-Te}}$ (\AA) \\
\hline
0 & 0.95 & 2.831 \\
2 & 1.07 & 2.822 \\
4 & 1.17 & 2.814 \\
5 & 1.21 & 2.811 \\
\end{tabular}
\end{ruledtabular}
\end{table}

\FloatBarrier

\section{Finite-temperature structural robustness}
\label{sec:aimd}

\begin{figure}[!ht]
\centering
\includegraphics[width=0.85\textwidth]{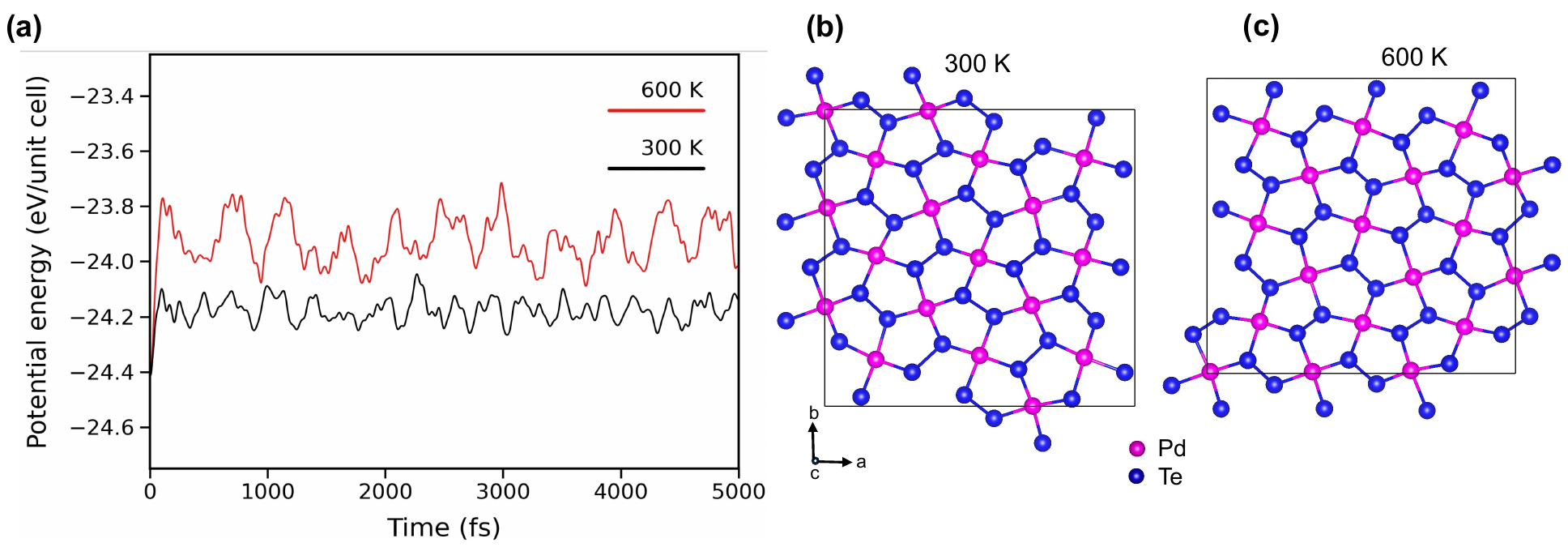}
\caption{AIMD potential-energy trajectories and final configurations of
pentagonal PdTe$_2$ at 300 and 600~K. (a) Potential energy per six-atom
$\mathrm{Pd}_2\mathrm{Te}_4$ unit cell as a function of time.
(b),(c) Top views of the final $3 \times 3 \times 1$ configurations at 300 and
600~K respectively.}
\label{fig:S1}
\end{figure}

\begin{figure}[!ht]
\centering
\includegraphics[width=0.85\textwidth]{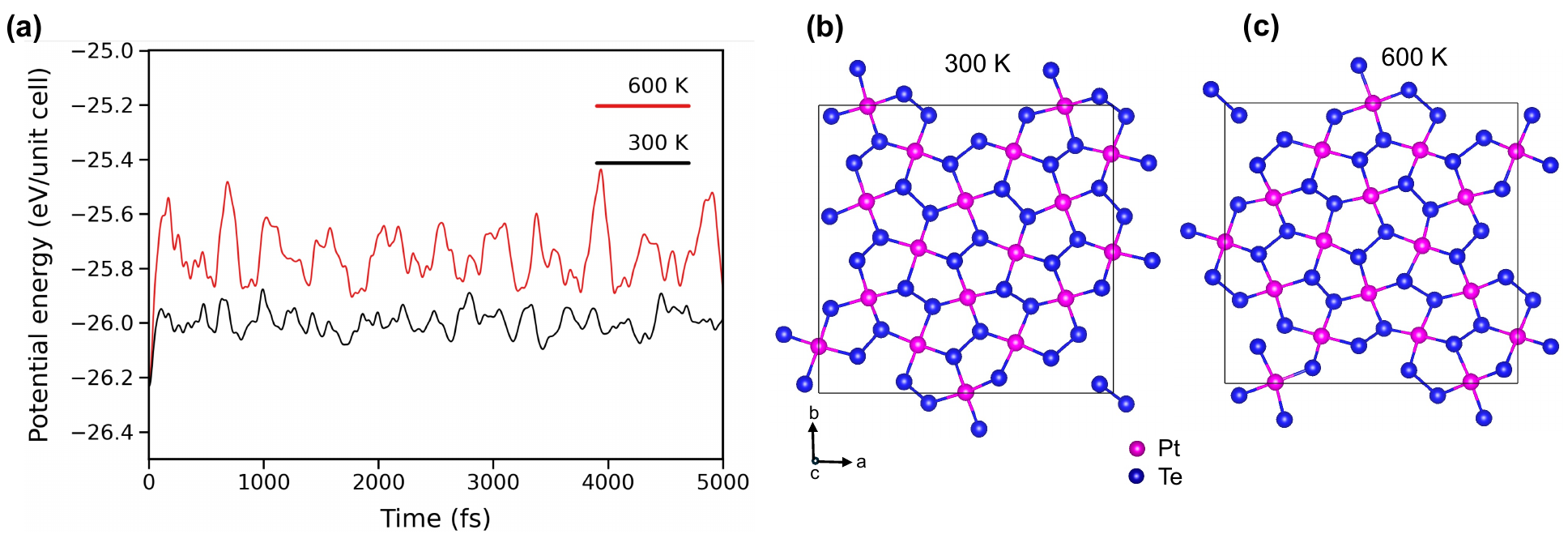}
\caption{AIMD potential-energy trajectories and final configurations of
pentagonal PtTe$_2$ at 300 and 600~K. Panels are as in
Fig.~\ref{fig:S1}.}
\label{fig:S2}
\end{figure}

\begin{figure}[!ht]
\centering
\includegraphics[width=0.85\textwidth]{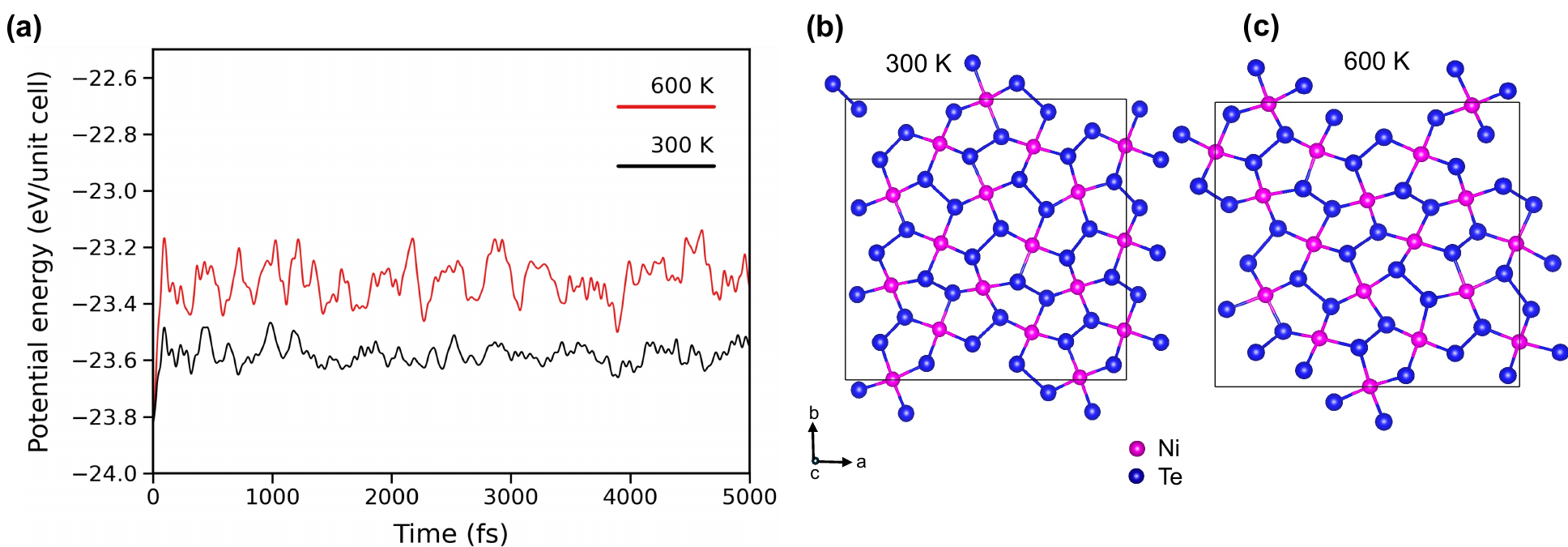}
\caption{AIMD potential-energy trajectories and final configurations of
pentagonal NiTe$_2$ at 300 and 600~K. Panels are as in
Fig.~\ref{fig:S1}.}
\label{fig:S3}
\end{figure}

Figures~\ref{fig:S1} to \ref{fig:S3} summarize the \textit{ab initio}
molecular dynamics results for pentagonal PdTe$_2$, PtTe$_2$ and NiTe$_2$
respectively. For each compound, panel (a) shows the potential energy as a
function of time during the 5~ps $NVT$ simulations at 300 and 600~K, while
panels (b) and (c) show top views of the corresponding final configurations.
The energies are reported per six-atom $X_2\mathrm{Te}_4$ pentagonal unit cell
and only the relative fluctuations within each trajectory are significant here.
The configurations represent periodic $3 \times 3 \times 1$ supercells. Once
each system has equilibrated at the set temperature, the potential energy
fluctuates about a constant mean with no systematic drift.

To characterise the dimer network throughout each trajectory rather than only
in the final configuration, we measured the Te--Te dimer separations in every
configuration recorded after equilibration. Each $3 \times 3 \times 1$
supercell contains eighteen Te--Te dimers. Table~\ref{tab:aimd} lists the mean
separation, averaged over all eighteen dimers and over the whole trajectory,
alongside the relaxed value of each compound. At 300~K the three means
reproduce the relaxed values, and at 600~K they increase by only 0.01 to
0.03~\AA, consistent with ordinary thermal expansion of the bond. The dimerized
bonding motif that underlies the semiconducting gap is therefore retained at
both temperatures.

\begin{table}[htbp]
\caption{Mean Te--Te dimer separation during the $NVT$ AIMD trajectories,
averaged over the eighteen dimers of the $3 \times 3 \times 1$ supercell and
over all configurations recorded after equilibration, compared with the relaxed
value of each compound.}
\label{tab:aimd}
\begin{ruledtabular}
\begin{tabular}{cccc}
Compound & $T$ (K) & $\langle d_{\mathrm{Te-Te}} \rangle$ (\AA) & Relaxed $d_{\mathrm{Te-Te}}$ (\AA) \\
\hline
\multirow{2}{*}{PdTe$_2$} & 300 & 2.81 & \multirow{2}{*}{2.81} \\
                          & 600 & 2.82 & \\
\hline
\multirow{2}{*}{PtTe$_2$} & 300 & 2.83 & \multirow{2}{*}{2.83} \\
                          & 600 & 2.84 & \\
\hline
\multirow{2}{*}{NiTe$_2$} & 300 & 2.84 & \multirow{2}{*}{2.84} \\
                          & 600 & 2.87 & \\
\end{tabular}
\end{ruledtabular}
\end{table}

\FloatBarrier
\clearpage

\section{Symmetry-enforced Dirac states}
\label{sec:dirac}

\begin{figure}[htbp]
\centering
\includegraphics[width=0.85\textwidth]{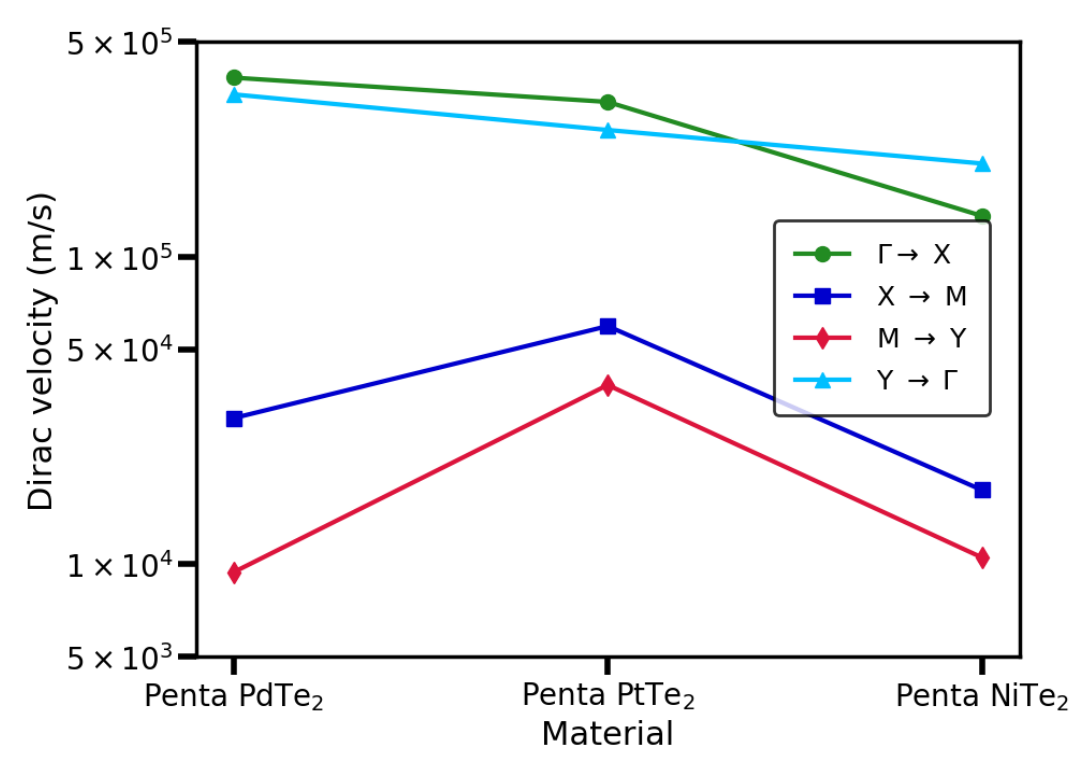}
\caption{Dirac velocities of the symmetry-enforced Dirac cones at the X and Y
points for pentagonal PdTe$_2$, PtTe$_2$ and NiTe$_2$. Velocities are extracted
from linear fits to the Dirac crossings along the high-symmetry directions
$\Gamma \rightarrow X$, $X \rightarrow M$, $M \rightarrow Y$ and
$Y \rightarrow \Gamma$. The vertical axis is plotted on a logarithmic scale to
allow comparison across the wide range of velocities. Directions toward
$\Gamma$ exhibit substantially larger velocities than those along the
zone-boundary directions toward $M$, highlighting the strong anisotropy of the
Dirac cones.}
\label{fig:S4}
\end{figure}

Figure~\ref{fig:S4} compares the Dirac velocities extracted at the X
and Y points of the three pentagonal ditellurides. Table~\ref{tab:coords} lists
the pristine and symmetry-broken atomic coordinates used to generate Fig.~3(c)
of the main text.

\begin{table}[htbp]
\caption{Cartesian atomic coordinates for pristine and distorted pentagonal
PdTe$_2$. The displacements break the nonsymmorphic $p2_1/b11$ layer-group
symmetry and gap the Dirac crossings, as shown in Fig.~3(c) of the main text.
All coordinates and displacements are in \AA.}
\label{tab:coords}
\begin{ruledtabular}
\begin{tabular}{lccc}
Atom & Pristine $(x,y,z)$ & Distorted $(x,y,z)$ & $|\Delta \mathbf{r}|$ \\
\hline
Pd1 & (3.2281, 0.0000, 10.0000) & (3.2620, 0.0262, 10.0550) & 0.070 \\
Pd2 & (0.0000, 3.0794, 10.0000) & (0.0210, 3.0501, 9.9550) & 0.058 \\
Te1 & (2.3759, 2.3454, 10.8379) & (2.4356, 2.4393, 10.9529) & 0.160 \\
Te2 & (5.6039, 0.7339, 10.8379) & (5.5587, 0.7555, 10.7529) & 0.099 \\
Te3 & (4.0803, 3.8133, 9.1621)  & (4.1110, 3.7455, 9.3171)  & 0.160 \\
Te4 & (0.8522, 5.4248, 9.1621)  & (0.7634, 5.4124, 9.0271)  & 0.099 \\
\end{tabular}
\end{ruledtabular}
\end{table}

\FloatBarrier
\clearpage

\section{Construction of the transformation pathways}
\label{sec:pathmethods}

Two independent pathways connecting the hexagonal and pentagonal structures of
PdTe$_2$ were constructed. Both use the same two endpoints, which were fully
relaxed with the production parameters of Sec.~II~A of the main text before
either pathway was built.

\subsection{Interpolation pathway}

The construction of the interpolation pathway is given in Sec.~II~B of the main
text. Two points deserve emphasis here.

First, the correspondence between the rectangular hexagonal supercell and the
pentagonal unit cell is not unique. The two inequivalent orientations of the
supercell relative to the pentagonal lattice, which differ by a rotation of
90 degrees, generate distinct pathways between the same pair of endpoint
structures. Both were constructed and the one with the lower energy profile at
every value of $\lambda$ was adopted. This is the correspondence in which the
long axis of the supercell maps onto the long axis of the pentagonal cell, so
that $a$ runs from 6.95 to 6.46~\AA{} and $b$ runs from 4.02 to 6.16~\AA{}
along the pathway.

Second, at each $\lambda$ the in-plane lattice vectors are held fixed at their
linearly interpolated values while the internal coordinates are relaxed. The
pathway therefore samples a restricted subspace of the full configuration
space, and the maximum of its energy profile is an upper bound on the
saddle-point energy for a concerted, cell-periodic transformation rather than
the saddle-point energy itself.

\subsection{Climbing-image solid-state nudged elastic band}

The minimum-energy path was obtained independently with the climbing-image
solid-state nudged elastic band method (CI-SSNEB)
\cite{Henkelman2000,Sheppard2012} as implemented in the VTST extension to VASP.
The band consisted of eleven images numbered 0 to 10, of which images 0 and 10
are the fully relaxed hexagonal $2 \times 1$ supercell and the fully relaxed
pentagonal structure and were held fixed. The nine interior images were
initialized from the interpolation pathway described above and were then
relaxed in the atomic and cell degrees of freedom simultaneously, with the cell
shape and volume free to change and the relative weighting of the cell and
atomic variables left at the default of the VTST implementation. A spring
constant of 5.0~eV/\AA$^2$ was applied between adjacent images and the
climbing-image variant was active throughout, so that the highest-energy image
converges onto the saddle point rather than onto an equally spaced position
along the band. The band was optimized with the FIRE algorithm of the VTST
toolkit, with a maximum step of 0.15~\AA, until the force on every image fell
below 0.01~eV/\AA.

The electronic-structure parameters were those used for the endpoint
relaxations, with a plane-wave cutoff of 520~eV, a $\Gamma$-centered
$12 \times 12 \times 1$ $k$-point mesh and an electronic convergence criterion
of $10^{-8}$~eV. Spin--orbit coupling was included in a fully noncollinear
treatment throughout the pathway optimization. The electronic band structures
shown for individual images in Fig.~\ref{fig:S6} were computed separately at
the converged image geometries, also with spin--orbit coupling.

Image 4, the fifth image of the band, carries the maximum of the converged
energy profile and is therefore the climbing image and the saddle point.

\subsection{Reaction coordinate}

Throughout this work the reaction coordinate is the interpolation parameter
$\lambda$ along the interpolation pathway, and the image index divided by ten
along the CI-SSNEB pathway, so that both run from 0 at the hexagonal structure
to 1 at the pentagonal structure. Because the elastic band distributes the
images at nearly uniform spacing, the second coordinate is approximately
proportional to the distance travelled in configuration space, which is what
makes the two pathways comparable on a common horizontal axis. The two are
nevertheless not equivalent parametrizations and agree only at the two
endpoints, so image $n$ should not be identified with $\lambda = n/10$.

The comparison of the two maxima does not depend on this correspondence.
Because both pathways connect the same two structures and the interpolation
pathway is confined to a subspace of the configuration space explored by the
CI-SSNEB band, the maximum of the interpolation profile bounds the CI-SSNEB
saddle from above for any parametrization of either path.

\FloatBarrier

\section{Evolution along the transformation pathways}
\label{sec:pathresults}

\begin{figure}[htbp]
\centering
\includegraphics[width=0.85\textwidth]{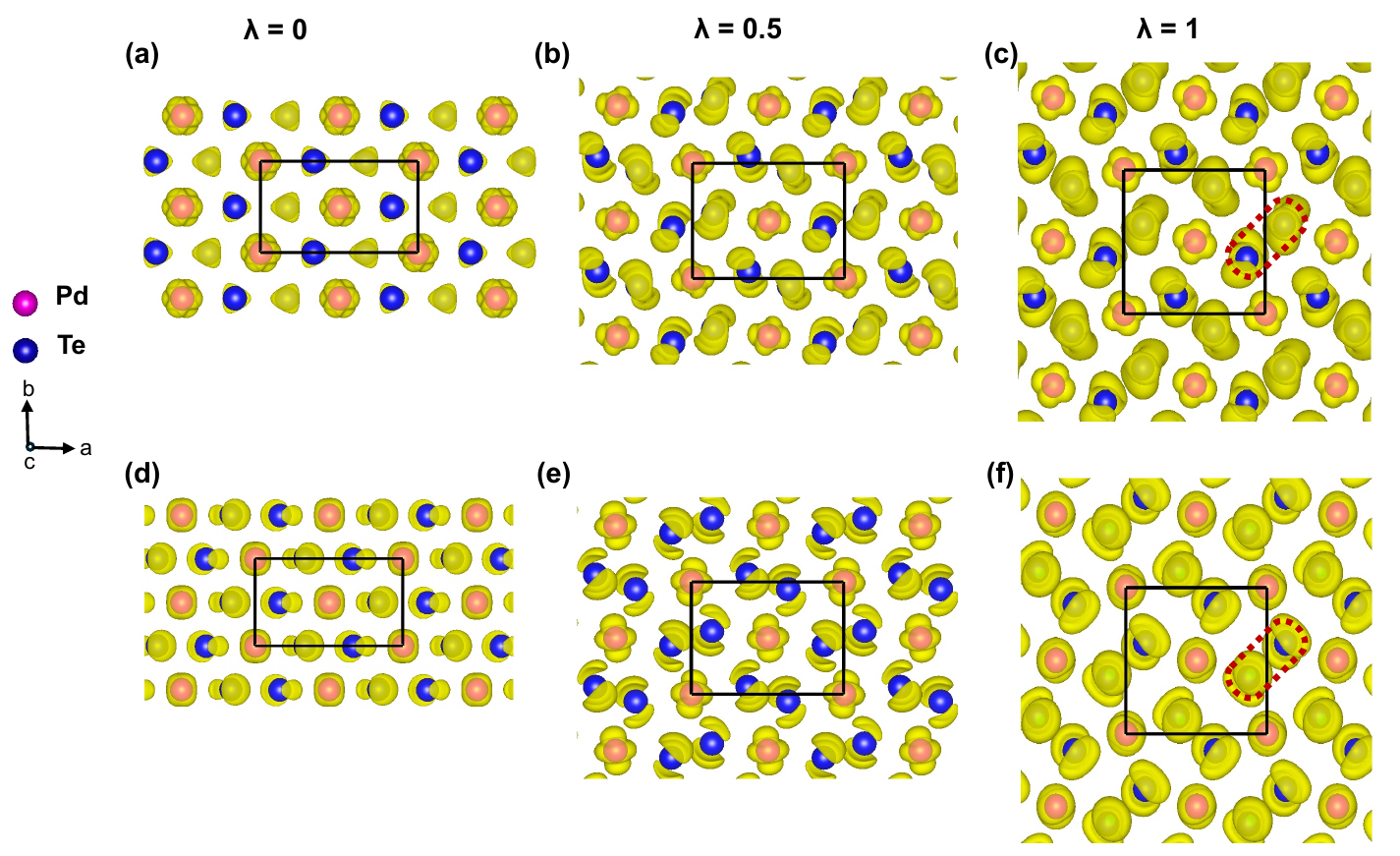}
\caption{Partial charge density isosurfaces of the conduction-band minimum
(CBM, top row) and valence-band maximum (VBM, bottom row) for monolayer
PdTe$_2$ along the hexagonal-to-pentagonal transformation pathway at
$\lambda = 0$ [(a),(d)], $\lambda = 0.5$ [(b),(e)] and $\lambda = 1$ [(c),(f)].
Yellow isosurfaces represent the band-edge charge density. As $\lambda$
increases, the charge density progressively develops enhanced localization on
neighboring Te pairs. At $\lambda = 1$, the VBM exhibits bond-centered charge
accumulation, while the CBM shows the corresponding antibonding counterpart,
consistent with Te--Te dimerization-driven bonding--antibonding splitting
responsible for opening the semiconducting gap. The isosurface values are
0.0049, 0.0040 and 0.0030 e/\AA$^3$ for $\lambda = 0$, 0.5 and 1 respectively.
The dashed red outlines in panels (c) and (f) highlight one of the fully formed
Te--Te dimers in the pentagonal phase.}
\label{fig:S5}
\end{figure}

\begin{figure*}[htbp]
\centering
\includegraphics[width=0.85\textwidth]{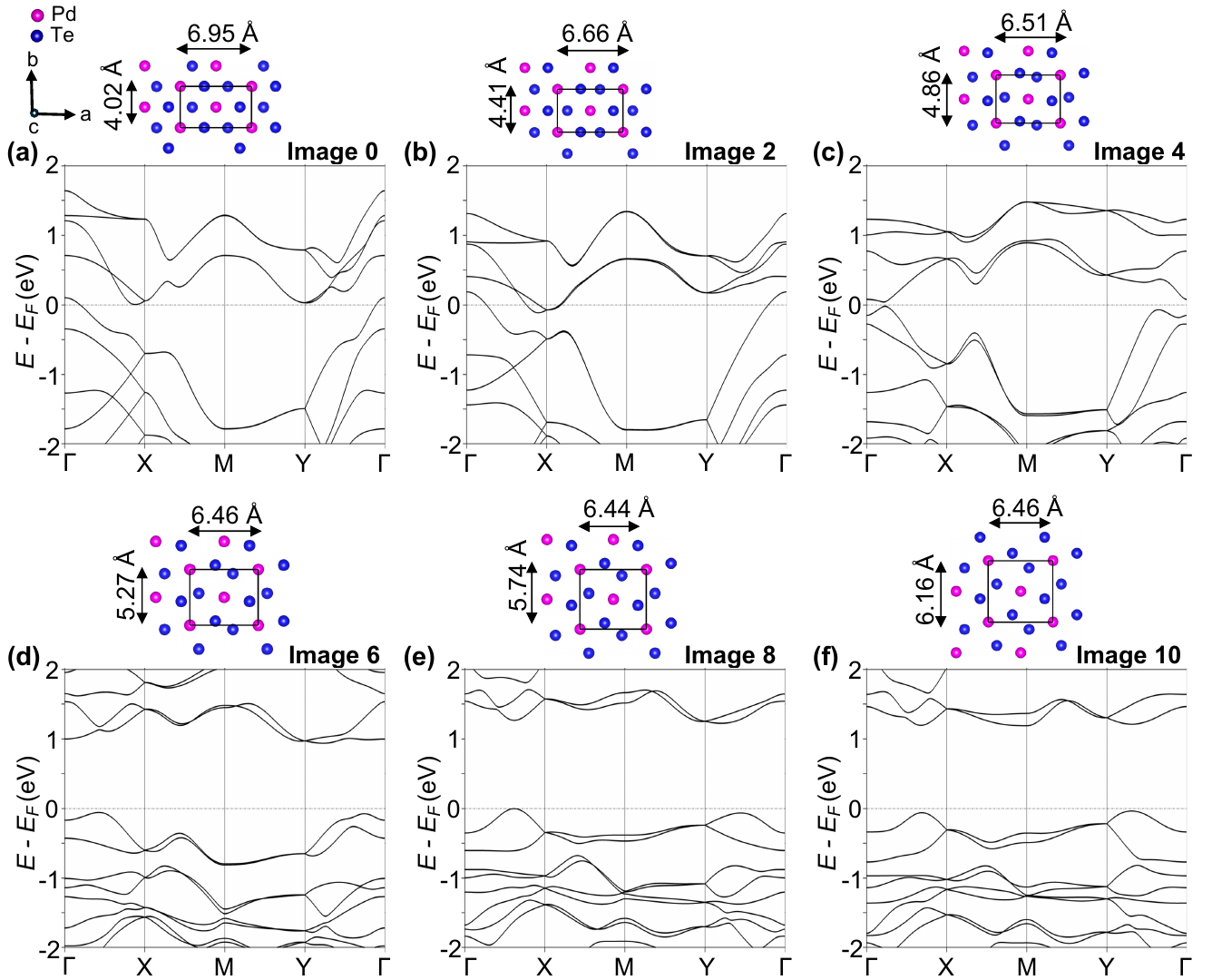}
\caption{Evolution of the crystal and electronic structures of PdTe$_2$ along
the climbing-image solid-state nudged elastic band (CI-SSNEB) pathway
connecting the hexagonal and pentagonal phases. (a)--(f) show every second
image of the eleven-image band, numbered 0 to 10 along the reaction coordinate.
Images 0 and 10 are the fully relaxed hexagonal and pentagonal endpoints and
were held fixed during the CI-SSNEB calculation, while the nine intermediate
images were converged along the pathway. Image 4 is the climbing image and
corresponds to the saddle point, which lies 0.357~eV per formula unit above the
hexagonal phase. The image index is not equivalent to the interpolation
parameter $\lambda$ of Fig.~4 of the main text, since the two pathways coincide
only at the endpoints, although both show the same qualitative evolution of the
electronic structure. For each image, the crystal structure (upper panel) and
the corresponding electronic band structure calculated with spin--orbit
coupling (lower panel) are shown. The in-plane lattice constants $a$ and $b$ of
the rectangular supercell are indicated in each structural panel. Magenta and
blue spheres denote Pd and Te atoms respectively. The system evolves from a
semimetallic state at image 0 to a semiconducting state at image 10, with the
band gap opening beyond image 4. Energies are referenced to the Fermi level.}
\label{fig:S6}
\end{figure*}

\begin{figure*}[htbp]
\centering
\includegraphics[width=\textwidth]{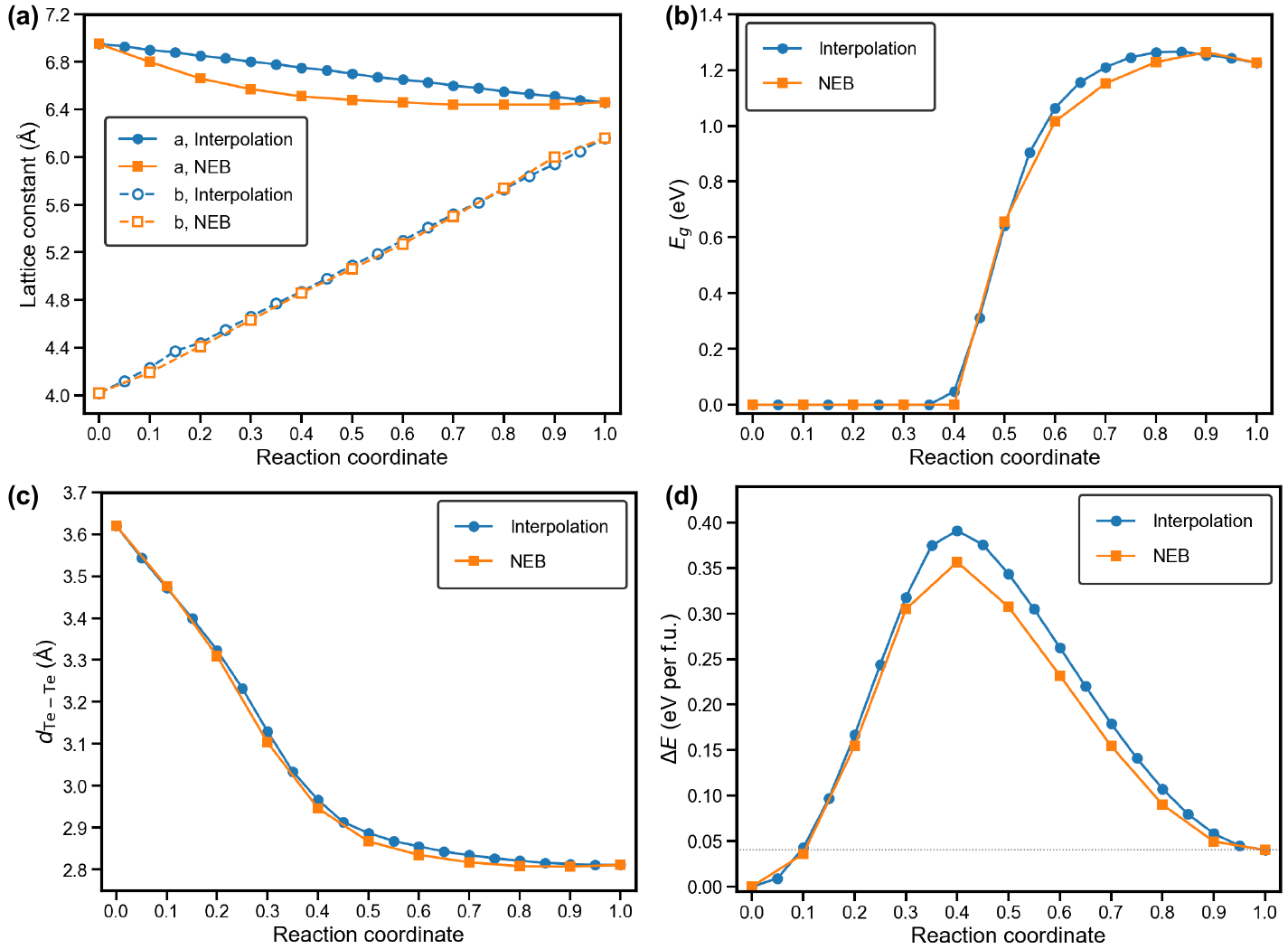}
\caption{Comparison of the interpolation and the climbing-image solid-state
nudged elastic band (CI-SSNEB) pathways for the hexagonal-to-pentagonal
transformation of monolayer PdTe$_2$, labelled Interpolation and NEB in the
panels. The reaction coordinate is the interpolation parameter $\lambda$ along
the interpolation pathway and the image index divided by ten along the
CI-SSNEB pathway, so that both run from 0 at the hexagonal structure to 1 at
the pentagonal structure. (a) In-plane lattice constants $a$ (filled symbols,
solid lines) and $b$ (open symbols, dashed lines) of the rectangular supercell.
(b) Band gap $E_g$ calculated with spin--orbit coupling. (c) Minimum Te--Te
separation. The dotted line marks the intrachain Te--Te bond length of
elemental trigonal Te, 2.835~\AA{} [Ref.~58 of the main text]. (d) Energy
relative to the hexagonal structure, per formula unit. The dotted line marks
the pentagonal structure at 0.04~eV per formula unit.}
\label{fig:S7}
\end{figure*}

Figure~\ref{fig:S5} shows the band-edge partial charge densities along the
interpolation pathway. Figure~\ref{fig:S6} shows the crystal and electronic
structures of alternate images of the CI-SSNEB band, and Fig.~\ref{fig:S7}
compares the two pathways directly.

The two pathways evolve almost identically in $b$, in the minimum Te--Te
separation and in the band gap, and differ principally in $a$. Along the
CI-SSNEB pathway $a$ contracts far more rapidly and has essentially reached its
final value by a reaction coordinate of 0.5, whereas along the interpolation
pathway it varies linearly by construction. The CI-SSNEB cell is therefore
about 4\% smaller in area through the middle of the transformation. The saddle
point of the CI-SSNEB pathway lies at 0.357~eV per formula unit, 8.8\% below
the maximum of 0.391~eV per formula unit reached along the interpolation
pathway, and the CI-SSNEB profile lies below the interpolation profile at every
intermediate image. The two endpoints are common to both calculations, and the
pentagonal structure lies 0.04~eV per formula unit above the hexagonal one, in
agreement with the zero-strain separation of Fig.~5(a) of the main text.

The electronic evolution is likewise reproduced. The gap remains closed up to a
reaction coordinate of 0.4 on both pathways. A residual gap of 48~meV is
already present at $\lambda = 0.40$ on the interpolation pathway, while image 4
of the CI-SSNEB band is still gapless, and the gap rises steeply thereafter in
both cases. Along the interpolation pathway the minimum Te--Te separation falls
below the intrachain Te--Te bond length of elemental trigonal Te, 2.835~\AA, at
$\lambda = 0.70$, at which point the gap has reached 99\% of its value in the
fully relaxed pentagonal structure. The CI-SSNEB pathway reaches the same
separation slightly earlier in the reaction coordinate, at image 6, consistent
with its more rapid contraction of $a$. Both pathways show the same weak
nonmonotonicity near the pentagonal end, with the gap passing through a shallow
maximum of 1.266~eV at $\lambda = 0.85$ and 1.264~eV at CI-SSNEB image 9 before
settling at 1.226~eV in the fully relaxed pentagonal structure. The dimer
length is already saturated in this region, so the 40~meV variation reflects
the residual evolution of the cell metric rather than a change in the dimerized
bonding motif. The close agreement between the two pathways in every structural
and electronic observable indicates that the interpolation pathway captures the
transformation mechanism.

\FloatBarrier
\clearpage

\section{Pentagonal sulfides and selenides}
\label{sec:chalcogenides}

\begin{figure}[htbp]
\centering
\includegraphics[width=0.85\textwidth]{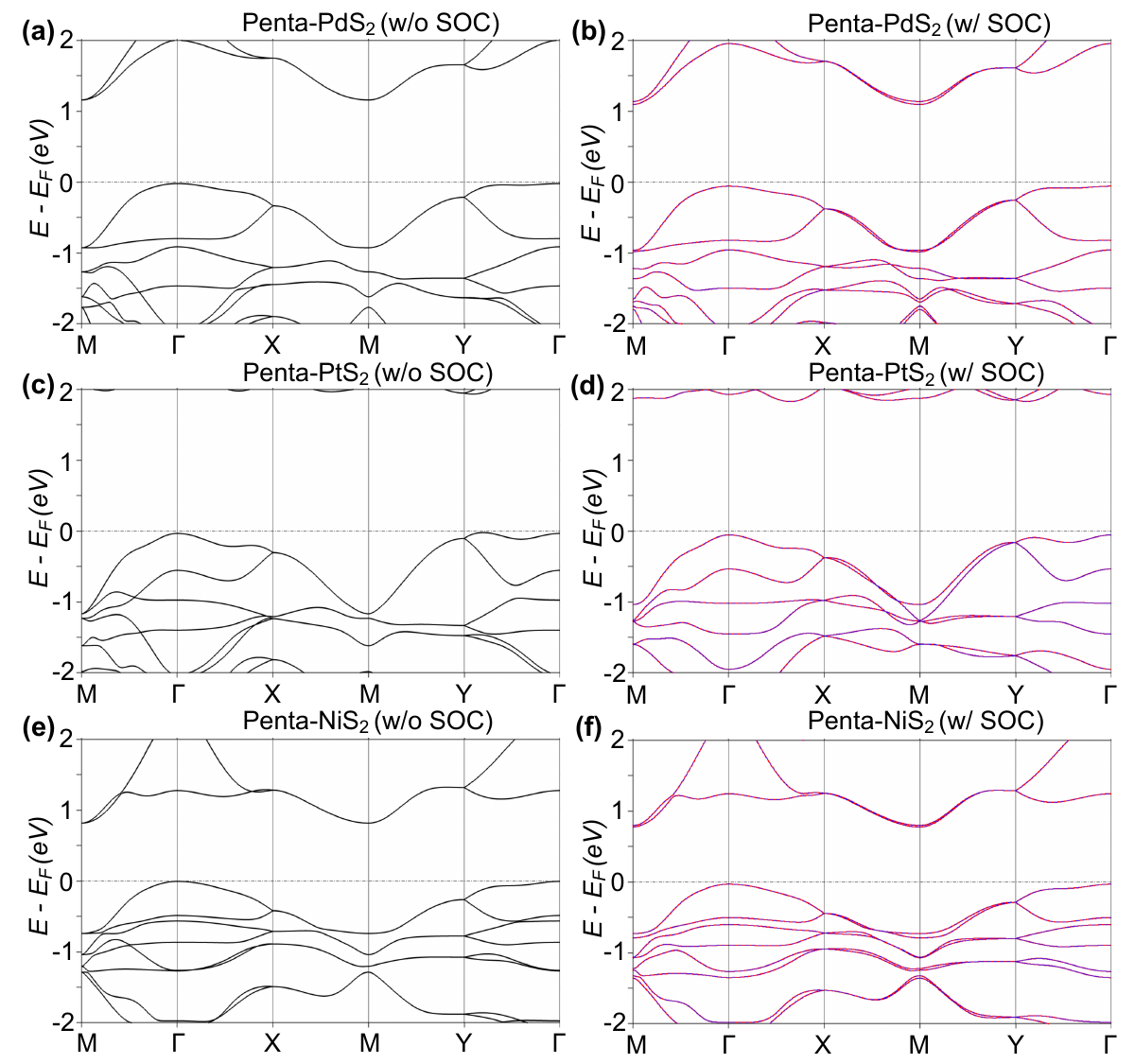}
\caption{Electronic band structures of pentagonal sulfides PdS$_2$, PtS$_2$ and
NiS$_2$ with and without spin--orbit coupling (SOC). Panels (a,b) correspond to
PdS$_2$, (c,d) to PtS$_2$ and (e,f) to NiS$_2$. As in the telluride compounds
discussed in the main text, the spinless band structures exhibit nonsymmorphic
symmetry-enforced boundary degeneracies along the X--M and M--Y directions,
while SOC lifts the boundary-line degeneracies and preserves fourfold Dirac
crossings at the Brillouin-zone boundary time-reversal invariant momenta X and
Y. The fourfold degeneracy at X and Y is enforced by the argument of
Sec.~III~C of the main text, which depends only on the layer group, inversion
and time reversal and therefore not on the strength of SOC or on the chemical
identity of the chalcogen. Energies are referenced to the Fermi level.}
\label{fig:S8}
\end{figure}

\begin{figure}[htbp]
\centering
\includegraphics[width=0.85\textwidth]{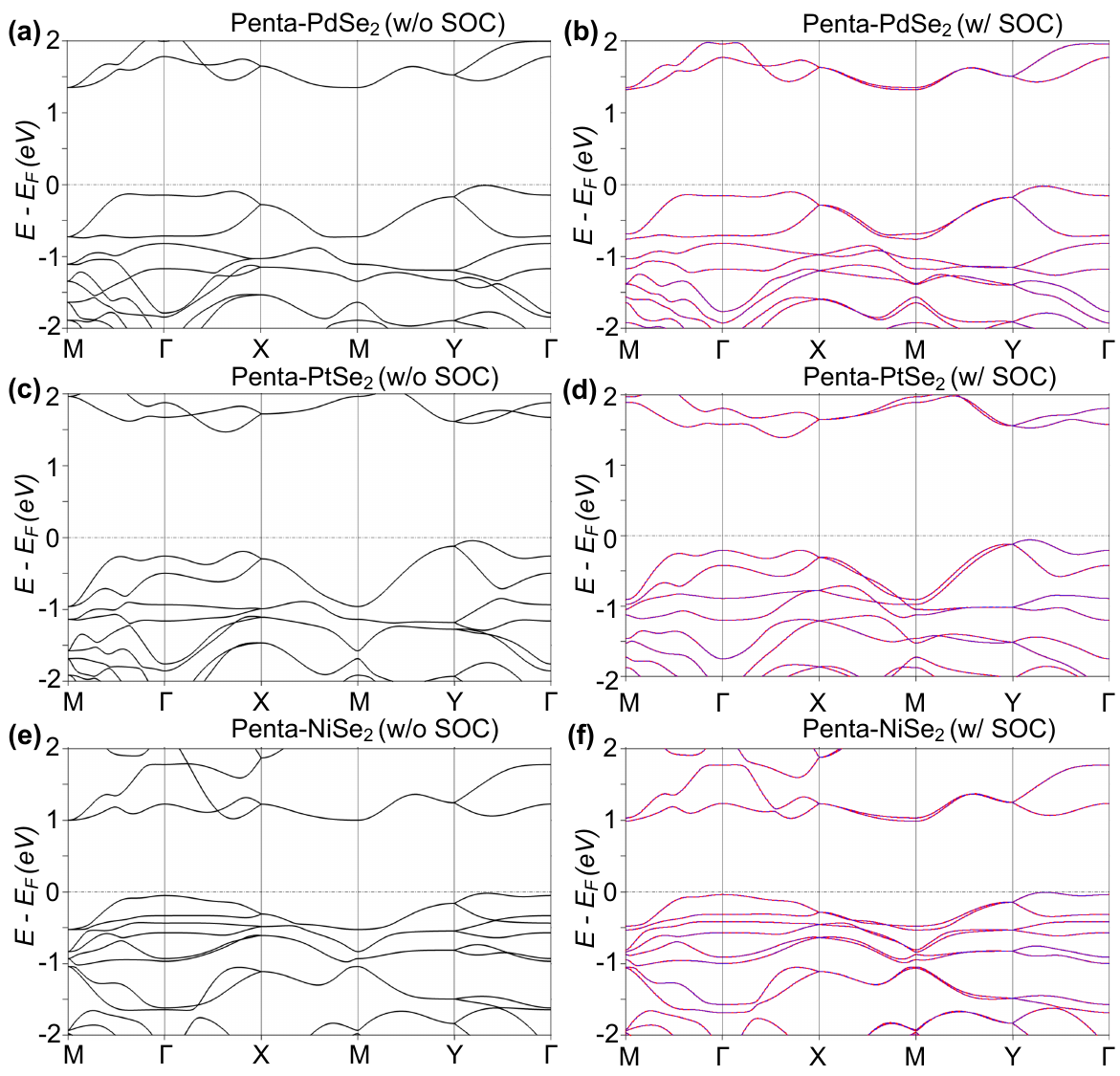}
\caption{Electronic band structures of pentagonal selenides PdSe$_2$, PtSe$_2$
and NiSe$_2$ with and without spin--orbit coupling (SOC). Panels (a,b)
correspond to PdSe$_2$, (c,d) to PtSe$_2$ and (e,f) to NiSe$_2$. As in the
telluride compounds presented in the main text, the spinless band structures
display nonsymmorphic symmetry-enforced boundary degeneracies along the X--M
and M--Y directions, while SOC removes the boundary-line degeneracies and
preserves symmetry-enforced fourfold Dirac crossings at the Brillouin-zone
boundary time-reversal invariant momenta X and Y. Energies are referenced to
the Fermi level.}
\label{fig:S9}
\end{figure}

Figures~\ref{fig:S8} and \ref{fig:S9} show the band structures of
the pentagonal sulfides and selenides, and Table~\ref{tab:dimer} lists the
relaxed chalcogen--chalcogen dimer length and band gap of all nine pentagonal
$XQ_2$ monolayers.

Two features of Table~\ref{tab:dimer} are relevant to the mechanism established
in the main text. First, the dimer length is determined almost entirely by the
chalcogen and is insensitive to the metal. Within each chalcogen the three
values span at most 0.04~\AA, whereas the separation changes by 0.70~\AA{}
between the sulfides and the tellurides, a ratio of about seventeen. This
metal-independence is characteristic of a genuine covalent $Q_2$ unit of fixed
bond order rather than a metal-mediated contact. For the tellurides the
calculated dimer length of 2.81 to 2.84~\AA{} lies within 0.03~\AA{} of the
intrachain Te--Te bond length of elemental trigonal Te, 2.835~\AA{} [Ref.~58 of
the main text]. Second, all nine compounds are semiconducting, in contrast to
the semimetallic or narrow-gap hexagonal $1T$ phases from which they derive.

\begin{table}[htbp]
\caption{Relaxed minimum chalcogen--chalcogen dimer length $d(Q\text{--}Q)$ and
band gap $E_g$ of the nine pentagonal $XQ_2$ monolayers. All structures are
fully relaxed and band gaps are computed including spin--orbit coupling at the
PBE level.}
\label{tab:dimer}
\begin{ruledtabular}
\begin{tabular}{lcc}
Material & $d(Q\text{--}Q)$ (\AA) & $E_g$ (eV) \\
\hline
PdS$_2$  & 2.11 & 1.16 \\
PtS$_2$  & 2.15 & 1.88 \\
NiS$_2$  & 2.13 & 0.81 \\
\hline
PdSe$_2$ & 2.42 & 1.34 \\
PtSe$_2$ & 2.46 & 1.45 \\
NiSe$_2$ & 2.45 & 0.99 \\
\hline
PdTe$_2$ & 2.81 & 1.23 \\
PtTe$_2$ & 2.83 & 1.32 \\
NiTe$_2$ & 2.84 & 0.92 \\
\end{tabular}
\end{ruledtabular}
\end{table}

\FloatBarrier
\clearpage

\section{Epitaxial constraint and in-plane strain}
\label{sec:strain}

\begin{figure}[htbp]
\centering
\includegraphics[width=0.6\textwidth]{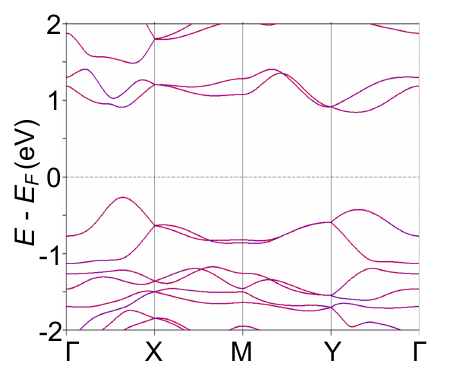}
\caption{Electronic band structure of pentagonal PdTe$_2$ constrained to the
$(\sqrt{5} \times \sqrt{5})R26.6^\circ$ surface cell of Pd(100),
$a = b = 6.15$~\AA, calculated with spin--orbit coupling. The fourfold Dirac
crossings at X and Y are preserved. Energies are referenced to the Fermi level.}
\label{fig:S10}
\end{figure}

The $(\sqrt{5} \times \sqrt{5})R26.6^\circ$ surface cell of Pd(100)
corresponds to a square lattice of 6.15~\AA. Relative to the free-standing
values of Table~I of the main text this amounts to a compression of 4.80\%
along $a$ and of 0.16\% along $b$, so the epitaxial constraint is strongly
anisotropic and is close to a uniaxial compression along $a$. The corresponding
change in cell area is 4.95\%. Table~\ref{tab:pdconstraint} compares the
free-standing monolayer with the same monolayer constrained to that cell, with
the internal coordinates fully relaxed, and Fig.~\ref{fig:S10} shows the band
structure of the constrained monolayer. The constrained structure retains the
$P2_1/c$ space group and the symmetry-enforced fourfold Dirac points, the
Te--Te dimer is retained at 2.786~\AA, and the band gap is 1.11~eV. The energy
cost of the epitaxial constraint is 68~meV per formula unit, which is small
compared with the barrier of 0.357~eV per formula unit separating the two
phases.

Table~\ref{tab:strain} reports the band gap of the free-standing pentagonal
PdTe$_2$ monolayer under biaxial, uniaxial and anisotropic in-plane strain of
up to $\pm 5$\%. Every strained structure retains the $P2_1/c$ space group and
the fourfold Dirac crossings at X and Y. This follows from symmetry rather than
from numerical accident, since a homogeneous strain along $a$ or $b$ leaves
every operation of the layer group invariant, so the anticommutation of
inversion and glide that enforces the fourfold degeneracy is preserved for such
a strain of any magnitude. The gap varies between 0.65 and 1.26~eV over the
range considered and does not close in any case, and the anisotropic strains,
in which one in-plane axis is stretched and the other compressed so that the
cell area is nearly unchanged, perturb the gap least.

\begin{table}[htbp]
\caption{Comparison of the free-standing pentagonal PdTe$_2$ monolayer with the
same monolayer constrained to the $(\sqrt{5} \times \sqrt{5})R26.6^\circ$
surface cell of Pd(100). In the constrained calculation the in-plane lattice
vectors were fixed and the internal coordinates were fully relaxed. Both
structures belong to the $P2_1/c$ space group and both host the
symmetry-enforced fourfold Dirac points at X and Y. Band gaps include
spin--orbit coupling.}
\label{tab:pdconstraint}
\begin{ruledtabular}
\begin{tabular}{lcccccc}
Structure & $a$ (\AA) & $b$ (\AA) & $E_{\mathrm{tot}}$ (eV) & $E_g$ (eV) & $d_{\mathrm{Te-Te}}$ (\AA) & Space group \\
\hline
Free-standing          & 6.46 & 6.16 & $-24.895$ & 1.23 & 2.811 & $P2_1/c$ \\
Constrained to Pd(100) & 6.15 & 6.15 & $-24.759$ & 1.11 & 2.786 & $P2_1/c$ \\
\end{tabular}
\end{ruledtabular}
\end{table}

\begin{table}[htbp]
\caption{Band gap of pentagonal PdTe$_2$ under in-plane strain. The internal
coordinates were relaxed at each imposed cell with all other parameters held at
their production values. Anisotropic strain stretches one in-plane axis and
compresses the other by the same amount, with positive $\varepsilon$ denoting
$a$ stretched and $b$ compressed and negative $\varepsilon$ the reverse. All
strained structures retain the $P2_1/c$ space group and the symmetry-enforced
fourfold Dirac crossings at X and Y. Band gaps include spin--orbit coupling.}
\label{tab:strain}
\begin{ruledtabular}
\begin{tabular}{lcc}
Strain type & $\varepsilon$ (\%) & $E_g$ (eV) \\
\hline
Unstrained         & 0    & 1.23 \\
\hline
Biaxial            & $-5$ & 1.07 \\
Biaxial            & $-3$ & 1.18 \\
Biaxial            & $+3$ & 0.87 \\
Biaxial            & $+5$ & 0.65 \\
\hline
Uniaxial along $a$ & $-5$ & 1.10 \\
Uniaxial along $a$ & $+5$ & 0.91 \\
\hline
Uniaxial along $b$ & $-5$ & 1.26 \\
Uniaxial along $b$ & $+5$ & 0.93 \\
\hline
Anisotropic        & $-5$ & 1.06 \\
Anisotropic        & $-3$ & 1.16 \\
Anisotropic        & $+3$ & 1.17 \\
Anisotropic        & $+5$ & 1.11 \\
\end{tabular}
\end{ruledtabular}
\end{table}

\FloatBarrier
\clearpage

\bibliographystyle{apsrev4-2}
\bibliography{PdTe2_paper}